\documentclass[fleqn,usenatbib]{mnras}

\usepackage{newtxtext,newtxmath}

\usepackage[T1]{fontenc}

\DeclareRobustCommand{\VAN}[3]{#2}
\let\VANthebibliography\thebibliography
\def\thebibliography{\DeclareRobustCommand{\VAN}[3]{##3}\VANthebibliography}

\usepackage{graphicx}	
\usepackage{amsmath}	

\newcommand{\hMpc}{h^{-1}{\rm Mpc}}
\newcommand{\hMsun}{h^{-1}{\rm M}_\odot}

\title[Improving the one-halo model for the galaxy-halo connection]{The MillenniumTNG Project: Refining the one-halo  model of red and blue galaxies at different redshifts}

\author[B. Hadzhiyska et al.]{%
Boryana Hadzhiyska,$^{1,2,3}$\thanks{E-mail: boryana.hadzhiyska@cfa.harvard.edu}
Lars Hernquist$^{1}$,
Daniel Eisenstein$^{1}$,
Ana Maria Delgado$^{1}$,
Sownak Bose$^{4}$,
\newauthor%
Rahul Kannan$^{1}$,
R\"udiger Pakmor$^{5}$,
Volker Springel$^{5}$,
Sergio Contreras$^{6}$,
Monica Barrera$^{5}$,
Fulvio Ferlito$^{5}$,
\newauthor%
C\'esar Hern\'andez-Aguayo$^{5,7}$,
Simon D. M. White$^{5}$,
and Carlos Frenk$^{4}$
\\%
\\%
$^{1}$Harvard-Smithsonian Center for Astrophysics, 60 Garden St, Cambridge, MA 02138, USA\\%
$^{2}$Miller Institute for Basic Research in Science, University of California, Berkeley, CA, 94720, USA\\%
$^{3}$Physics Division, Lawrence Berkeley National Laboratory, Berkeley, CA 94720\\
$^{4}$Institute for Computational Cosmology, Department of Physics, Durham University, South Road, Durham, DH1 3LE, UK\\%
$^{5}$Max-Planck-Institut f\"{u}r Astrophysik, Karl-Schwarzschild-Str. 1, 85748, Garching, Germany\\%
$^{6}$Donostia International Physics Center (DIPC), Donostia-San Sebastian, Spain\\%
$^{7}$Excellence Cluster ORIGINS, Boltzmannstrasse 2, 85748 Garching, Germany\\
}

\date{Accepted XXX. Received YYY; in original form ZZZ}

\pubyear{2022}

\begin{document}
\label{firstpage}
\pagerange{\pageref{firstpage}--\pageref{lastpage}}
\maketitle
 
\begin{abstract}
Luminous red galaxies (LRGs) and blue star-forming emission-line galaxies (ELGs) are key tracers of large-scale structure used by cosmological surveys. Theoretical predictions for such data are often done via simplistic models for the galaxy-halo connection. In this work, we use the large, high-fidelity hydrodynamical simulation of the MillenniumTNG project (MTNG) to inform a new phenomenological approach for obtaining an accurate and flexible galaxy-halo model on small scales. Our aim is to study LRGs and ELGs at two distinct epochs, $z = 1$ and $z = 0$, and recover their clustering down to very small scales, $r \sim 0.1 \ \hMpc$, i.e. the one-halo regime,  while a companion paper extends this to a two-halo model for larger distances.   The occupation statistics of ELGs in MTNG inform us that: (1) the satellite occupations exhibit a slightly super-Poisson distribution, contrary to commonly made assumptions, and (2) that haloes containing at least one ELG satellite are twice as likely to host a central ELG. We propose simple recipes for modeling these effects, each of which calls for the addition of a single free parameter to simpler halo occupation models. To construct a reliable satellite population model, we explore the LRG and ELG satellite radial and velocity distributions and compare them with those of subhalos and particles in the simulation. We find that ELGs are anisotropically distributed within halos, which together with our occupation results provides strong evidence for cooperative galaxy formation (manifesting itself as one-halo galaxy conformity); i.e.~galaxies with similar properties form in close proximity to each other.  Our refined  galaxy-halo model represents a useful improvement of commonly used analysis tools and thus can be of help to increase the constraining power of large-scale structure surveys.
\end{abstract}


\section{Introduction}
\label{sec:intro}

With the advent of large photometric and spectroscopic galaxy surveys such as the Dark Energy Survey \citep[DES,][]{1708.01530}, the Dark Energy Spectroscopic Instrument  \citep[DESI,][]{2016arXiv161100036D}, and the upcoming Vera Rubin Observatory \citep[\textit{Rubin,}][]{0912.0201} and Nancy Grace Roman Space Telescope \citep[\textit{Roman,}][]{2015arXiv150303757S}, the field of large-scale structure will enjoy an unprecedented improvement in the precision of galaxy clustering measurements, providing an extraordinary opportunity to refine our cosmological paradigm and expand our understanding of galaxy formation. One standard approach in galaxy surveys is to extract cosmological information from large-scale galaxy clustering by, e.g., using the baryon acoustic oscillations (BAO) as a `standard ruler' \citep{2005ApJ...633..560E}. But upcoming surveys will also measure the small-scale clustering signal with exquisite precision, allowing an alternative path to unravel long-standing mysteries pertaining to the nature of dark matter, dark energy, gravity, neutrinos, and the initial conditions of the Universe.

However, galaxy clustering on small scales is influenced by poorly understood baryonic processes, which makes the task of modeling this regime very challenging. Purely analytical models such as effective field theory \citep[see][for a review]{1406.7843}, in which one adopts a perturbative expansion to model both the non-linear dark matter structure growth and the galaxy tracing, have the downside of being limited to the linear and quasi-linear regimes. But since galaxies form and evolve in dark-matter haloes \citep{1978MNRAS.183..341W}, a more accurate and practical approach to determining the clustering of galaxies on all scales can be obtained by modeling the clustering of haloes jointly with the link between galaxies and their halo hosts, known as the galaxy-halo connection.

While numerical $N$-body simulations track the evolution of dark matter under the influence of gravity and are able to accurately predict the halo clustering on small scales, they remain agnostic about the physics of galaxies. Thus, on their own, they cannot provide us with knowledge about the crucial galaxy-halo link for connecting theory with observations. Cosmological, full-physics hydrodynamical simulations of galaxy formation \citep[e.g.,][]{2015MNRAS.446..521S,2019ComAC...6....2N}, on the other hand, simulate the dark matter component along with the gas and stars. In hydrodynamical simulations, the baryonic and galaxy processes are tracked by a combination of fluid equations and subgrid models. As a result of their complexity, this type of simulation is  prohibitively expensive to run in the large volumes needed for performing the observational analysis of upcoming surveys. However, smaller simulations can nevertheless provide invaluable insight into how and where galaxies form in relation to dark-matter hosts. Such insights have the potential of being highly beneficial for studying cosmology and galaxy formation from observations.

Only recently hydrodynamical simulations have become sufficiently large to be utilized in large-scale structure analysis \citep[e.g.,][]{2017MNRAS.465.2936M, 2018MNRAS.475..676S}. But until now, the  most well-resolved of these simulations have yielded only $\sim$10,000 galaxies at the typically expected number densities of current surveys ($\lesssim 10^{-3} \ [\hMpc]^{-3}$) due to the still limited volume, leading to substantial statistical uncertainties in measured quantities. The new cosmological full-physics MillenniumTNG (MTNG) box offers a significant improvement over previous cosmological hydrodynamical simulations in this regard. With a box size of $500\,h^{-1}{\rm Mpc}$ and $2\times 4230^3$ resolution elements, it not only offers a factor of $\sim$15 larger volume than TNG300-1, the largest simulation from the IllustrisTNG suite, it also maintains the high resolution of the IllustrisTNG suite and its successful physics model \citep[for a review of the IllustrisTNG project, see][]{2019ComAC...6....2N}. This provides sufficient statistics to study the galaxy samples targeted by near-future galaxy surveys (in particular emission-line galaxies and luminous red galaxies), making MTNG an ideal testing ground for developing and validating galaxy-halo tools for the pipeline analysis of these experiments.

One of the standard methods for analyzing small-scale galaxy clustering data from cosmological surveys involves equipping dark-matter-only simulations spread over different cosmologies with some `galaxy-painting' technique that allows a statistical comparison between theoretical predictions of the galaxy distribution with the observed galaxy catalogue. Typically, it is preferable to adopt computationally inexpensive methods that can produce the large number of mock catalogues necessary for cosmological inference. Empirical models such as the halo occupation distribution model \citep[HOD,][]{2002ApJ...575..587B, 2002PhR...372....1C,  2004MNRAS.350.1153Y, 2005ApJ...633..791Z} and subhalo abundance matching \citep[SHAM,][]{2006ApJ...647..201C, 2010ApJ...717..379B, 2014ApJ...783..118R, 2016MNRAS.459.3040G, 2016MNRAS.460.3100C} offer a simple and computationally inexpensive approach to modelling galaxy clustering by characterizing the relation between galaxies and host (sub)haloes. The HOD prescription dictates the average number of central and satellite galaxies in a given  halo as a function of only the halo mass, while SHAM connects galaxies to dark matter subhaloes using a monotonic relation between galaxy luminosity (or stellar mass) and subhalo mass (or maximum circular velocity). Both methods come with well-known shortcomings when fitting observational data \citep[e.g.,][]{Norberg, Zehavi, 2017MNRAS.467.3024L, 2019arXiv190705424Z}.  It is crucial to address and understand these limitations before applying these methods to galaxy surveys for the extraction of cosmological information.

In this paper, which is part of a set of introductory studies of the MillenniumTNG (MTNG) project, we address the question of how to best model the occupations and satellite distributions of haloes based on the currently largest hydrodynamical simulation of galaxy formation, the large-volume full-physics run of the MTNG simulations. An overview of the full simulation suite of MillenniumTNG and an analysis of matter clustering and halo statistics is given in \citet{Aguayo2022}, while \citet{Pakmor2022} provide a detailed description of the hydro simulation together with an examination of the properties of its galaxy clusters. Further introductory papers present analyses of high-redshift galaxies \citep{Kannan2022},  weak gravitational lensing \citep{Ferlito2022}, intrinsic alignment \citep{Delgado2022}, galaxy clustering \citep{Bose2022}, and semi-analytic galaxies on the  lightcone \citep{Barrera2022}.

In the present work, we test common assumptions about the halo occupation distribution such as the presumed Poisson distribution of satellites, and the independence of central and satellite occupations. We also study in detail the radial profiles and velocity distributions for the various galaxy samples, and investigate the supposed isotropy of the satellite distribution. We propose a simple and intuitive model for determining the occupation numbers of haloes and distribution of satellites, designed for creating realistic mock catalogues that can be used in observational analysis. We test its ability to reproduce statistics of the galaxy distribution in MTNG on small scales, $0.1 \lesssim r \lesssim 1 \ \hMpc$, i.e. the so-called one-halo regime. In a follow-up paper \citep{Hadzhiyska2022b}, we extend our analysis to larger scales and study assembly bias and the two-halo term.

The structure of this paper is as follows. In Section~\ref{sec:meth}, we concisely describe the MillenniumTNG simulation suite and the methods we adopt for selecting galaxies. In Section~\ref{sec:gal_pop}, we introduce our model of the halo occupations and satellite distributions.  Finally, in Section~\ref{sec:conc}, we summarize and discuss our findings.

\section{Methods}
\label{sec:meth}
 
\subsection{The MillenniumTNG simulations}
\label{sec:mtng}

The MillenniumTNG project aims to provide accurate numerical predictions in sufficiently large volumes to improve the link between non-linear galaxy formation and the evolution of large-scale structure. The full simulation suite consists of several hydrodynamical and $N$-body simulations at various resolutions and box sizes. In this study, we employ the largest available full-physics box, \textsc{MTNG740}, and its dark-matter-only (DMO) counterpart, containing $2 \times 4320^3$ and $4320^3$ resolution elements, respectively, in a comoving volume of $(0.5\,h^{-1} {\rm Gpc})^3$. For MTNG, this yields a mass resolution in the baryons of $2.1 \times 10^7 \hMsun$ while that of the dark matter is $1.1 \times 10^8 \hMsun$. Throughout this paper, we refer to these calculations as MTNG and MTNG-DMO.

MTNG uses the same physics model and cosmology as IllustrisTNG \citep{2017MNRAS.465.3291W,2018MNRAS.473.4077P,2018MNRAS.475..648P,2018MNRAS.475..624N,2018MNRAS.477.1206N,2018MNRAS.480.5113M,2018MNRAS.475..676S, 2019MNRAS.490.3234N, 2019MNRAS.490.3196P}, and its resolution is comparable to but slightly lower than that of the largest IllustrisTNG box, TNG300-1. MTNG also uses the same hydrodynamical moving-mesh code \textsc{AREPO} \citep[][]{2010MNRAS.401..791S}, a main feature of which is the use of a Voronoi tessellation for the construction of the computational mesh. \textsc{AREPO} employs a finite-volume method to solve the hyperbolic inviscid Euler equations on a moving, unstructured mesh that is able to follow the flow in a quasi-Lagrangian fashion without imposing any preferred directions. The dark matter in \textsc{AREPO} is treated through the usual particle-based discretization as an $N$-body system. The self-gravity of the gas and the dark matter is computed with a hierarchical multipole approximation, a so-called tree-algorithm, coupled to a particle mesh like approach to compute the long-range gravitational field with Fourier techniques \citep{Springel2005}. Haloes (groups) in MTNG are identified by applying the standard friends-of-friends \citep[FoF,][]{1985ApJ...292..371D} algorithm to the dark matter particles, adopting a linking length of $b = 0.2$ (in units of the mean interparticle distance). Gravitationally bound substructures in halos are identified with the {\small SUBFIND-HBT} algorithm described in \citet{Springel2021}. (For a full technical exposition of the MillenniumTNG simulations and a description of the available outputs we refer the reader to \citet{Aguayo2022} and \citet{Pakmor2022}.)

Throughout the text, we refer to the ``virial'' mass and radius of a halo as
the total mass enclosed in a sphere around the halo center\footnote{Taken as the location of the minimum of the gravitational potential of the largest subhalo in the group.} whose mean density is $\Delta_c$ times the critical density of the Universe. $\Delta_c$ is derived from the generalized solution of the collapse of a spherical top-hat perturbation in a low-density universe, and is well fit by the polynomial function \citep{1998ApJ...495...80B}:
\begin{equation}
  \Delta_c(z) = 18 \pi^2 + 82 x - 39 x^2 \, ,
\end{equation} 
where $x = \Omega_m(z) - 1$, and $\Omega_m(z)$ is the matter energy density at redshift $z$.

\subsection{Galaxy populations}
\label{sec:gal_sel}

The main targets of current and near-future galaxy redshift surveys such as DESI, \textit{Euclid}, and \textit{Roman} at $z \lesssim 2$ will be emission-line (ELGs) and luminous red galaxies (LRGs), with expected number densities roughly varying in the range $10^{-4} \lesssim n_{\rm gal} / (\hMpc)^{-3}\lesssim 10^{-3} $. In this work, we extract LRGs and ELGs at redshifts $z = [0.0, 1.0]$ with two number densities $n_{\rm gal} = [7.0 \times 10^{-4}, 2.0 \times 10^{-3}]  (\hMpc)^{-3}$, corresponding to  $N_{\rm gal} = [87 \ 800, 250 \ 000]$ galaxies, with the following procedure:
\begin{itemize}
    \item \textbf{ELGs} are selected by applying a stellar mass and a specific star formation rate (sSFR) cut to the subhalos in MTNG. At $z = 0$ and $z=1$, the corresponding minimum stellar masses are ${M_\ast} = 4.9 \times 10^9$ and $8.3 \times 10^9 \ \hMsun$, and the minimum sSFR thresholds are ${\rm sSFR} = 2.9 \times 10^{-10}$ and $8.2 \times 10^{-10} \, h\,{\rm yr}^{-1}$ for $n_{\rm gal} = 7.0 \times 10^{-4} \ [\hMpc]^{-3}$, while for $\ n_{\rm gal} = 2.0 \times 10^{-3} \ [\hMpc]^{-3}$, they are ${M_\ast} = 4.8 \times 10^9$ and $6 \times 10^9 \ \hMsun$, and ${\rm sSFR} = 2.0 \times 10^{-10}$ and $6.0 \times 10^{-10} \, h\,{\rm yr}^{-1}$, respectively. This selection of ELGs is based on \citet{2021MNRAS.502.3599H}, who find that the colour-selected ELG sample is congruous to one selected by sSFR-stellar mass. 
    \item \textbf{LRGs} are selected by applying a stellar mass cut to the subhalos in MTNG. Additionally, we impose a  maximum sSFR threshold matching that of the ELGs for each corresponding sample to ensure that there is no overlap between the LRGs and ELGs at a given redshift and number density. Moreover, making a sSFR selection in addition to a stellar mass one ensures that we choose the most massive \textit{quenched} galaxies, akin to how observational targets are selected. At $z = 0$ and $z=1$, the corresponding minimum stellar masses are ${M_\ast} = 1.1 \times 10^{11}$ and $7.3 \times 10^{10} \ \hMsun$ for $n_{\rm gal} = 7.0 \times 10^{-4} \ [\hMpc]^{-3}$, while for $n_{\rm gal} = 2.0 \times 10^{-3} \ [\hMpc]^{-3}$, they are ${M_\ast} = 4.8 \times 10^{10}$ and $2.8 \times 10^{10} \ \hMsun$, respectively.
    \end{itemize}
    
Throughout the following, we refer to the two number densities, $n_{\rm gal} = [7.0 \times 10^{-4}, 2.0 \times 10^{-3}] \ (\hMpc)^{-3}$, as ``low'' and ``high'' for short.
The two-point correlation functions used in this paper are computed using the natural estimator via the \textsc{Corrfunc} package \citep{2020MNRAS.491.3022S}.

\section{Results}
\label{sec:gal_pop}

In this section, we consider common assumptions about the occupation characteristics of satellite galaxies and propose minimal modifications to well-established recipes, designed to make them more flexible and accurate in order to better reproduce the clustering statistics of galaxies in MTNG. 

The two-point correlation function of a galaxy catalog receives contributions from the so-called one- and two-halo terms. The one-halo term is determined by the distribution of satellites within a halo, which in real space, is typically confined to scales $\lesssim 1 \ \hMpc$. In redshift space, however, the spatial and velocity distributions of satellites can affect the clustering on appreciably larger scales. Moreover, since the peculiar velocities of galaxies are sensitive to the large-scale gravitational field, they can be used to test modified gravity and deviations from $\Lambda$CDM in observational analyses. Thus, providing accurate predictions for the transition regime between the one- and two-halo terms is an important task for small-scale and weak lensing analyses.

\subsection{Halo occupation model}
\label{sec:halo_model}

\subsubsection{Mean halo occupation distribution}
\label{sec:hod}

As upcoming data sets drive the demands for large-volume simulations and multiple mock realizations \citep[][]{2008RSPTA.366.4381B}, simplicity and computational efficiency are highly desirable when generating mock galaxy catalogues. The halo occupation distribution (HOD) statistic provides a simple relationship between the halo catalogues output of an $N$-body simulation and the galaxies we find in them. As such, it can be used to produce a large number of mock galaxy catalogues relatively quickly and inexpensively.

The HOD shape of red (LRG-like) and magnitude-limited samples has been studied thoroughly in the literature, and has been shown to be well-approximated by the empirical formula given in \citet{Zheng:2004id}, according to which the mean halo occupation of centrals, $\langle N_{{\rm cen}} \rangle$, and satellites, $\langle N_{{\rm sat}} \rangle$, as a function of mass, $M$, is: 
\begin{equation}\label{eq:ncen}
    \langle N_{{\rm cen}} (M) \rangle = \frac{1}{2} \left[ 1 + {\rm erf} \left( \frac{\log M-\log M_{{\rm min}}}{\sigma_{{\log M}}} \right) \right] ,
\end{equation}
\begin{equation}\label{eq:nsat}
    \langle N_{{\rm sat}} (M)\rangle = \left( \frac{M-M_{{\rm cut}}}{M_1} \right)^\alpha \, ,
\end{equation}
where $M_{\rm min}$ is the characteristic minimum mass of halos that host central galaxies, $\sigma_{\log M}$ is the width of this transition, $M_{{\rm cut}}$ is the characteristic cut-off scale for hosting satellites, $M_1$ is a normalization factor, and $\alpha$ is a power-law slope. 

In the case of ELGs, the standard formula of Eq.~\eqref{eq:ncen} does not provide a good fit for the central occupations, as according to Eq.~\eqref{eq:ncen}, the mean occupation of centrals flattens to 1 for large halo masses \citep[see e.g.,][]{2020arXiv200709012A}. However, since the central galaxies of massive haloes are typically quenched and red, they do not exhibit strong emission lines and are thus unlikely to make it into an ELG-targeting sample. Better-suited for the ELG occupations is the High Mass Quenched (HMQ) model, proposed in \citet{2020MNRAS.497..581A}, which expresses the mean central occupation as:
\begin{align}
    \left< N_{\rm cen}\right>(M) &=  2 A \phi(M) \Phi(\gamma M)  + & \nonumber \\  
    \frac{1}{2Q} & \left[1+{\rm erf}\left(\frac{\log_{10}{M}-\log{M_{\rm min}}}{0.01}\right) \right],  \label{eq:NHMQ}\\
\phi(x) &=\mathcal{N}(\log{ M_{\rm min}},\sigma_M), \label{eq:NHMQ-phi}\\
\Phi(x) &= \int_{-\infty}^x \phi(t) \, \mathrm{d}t = \frac{1}{2} \left[ 1+{\rm erf} \left(\frac{x}{\sqrt{2}} \right) \right], \label{eq:NHMQ-Phi}\\
A &=\frac{p_{\rm max}  -1/Q }{\max[2\phi(x)\Phi(\gamma x)]},
\label{eq:NHMQ-A}
\end{align}
where the new parameters $p_{\rm max}$ and $Q$ control the amplitude and quenching of ELG central occupations, respectively. The occupation statistics of the satellites is assumed to obey the standard functional form of Eq.~\eqref{eq:nsat}. (For more details on the interpretation of the various parameters, see \citet{2020MNRAS.497..581A}.) The model seems to qualitatively match the shape of the ELG HOD predicted by IllustrisTNG \citep[][]{2021MNRAS.502.3599H}.

\begin{figure}
\centering  
\includegraphics[width=0.48\textwidth]{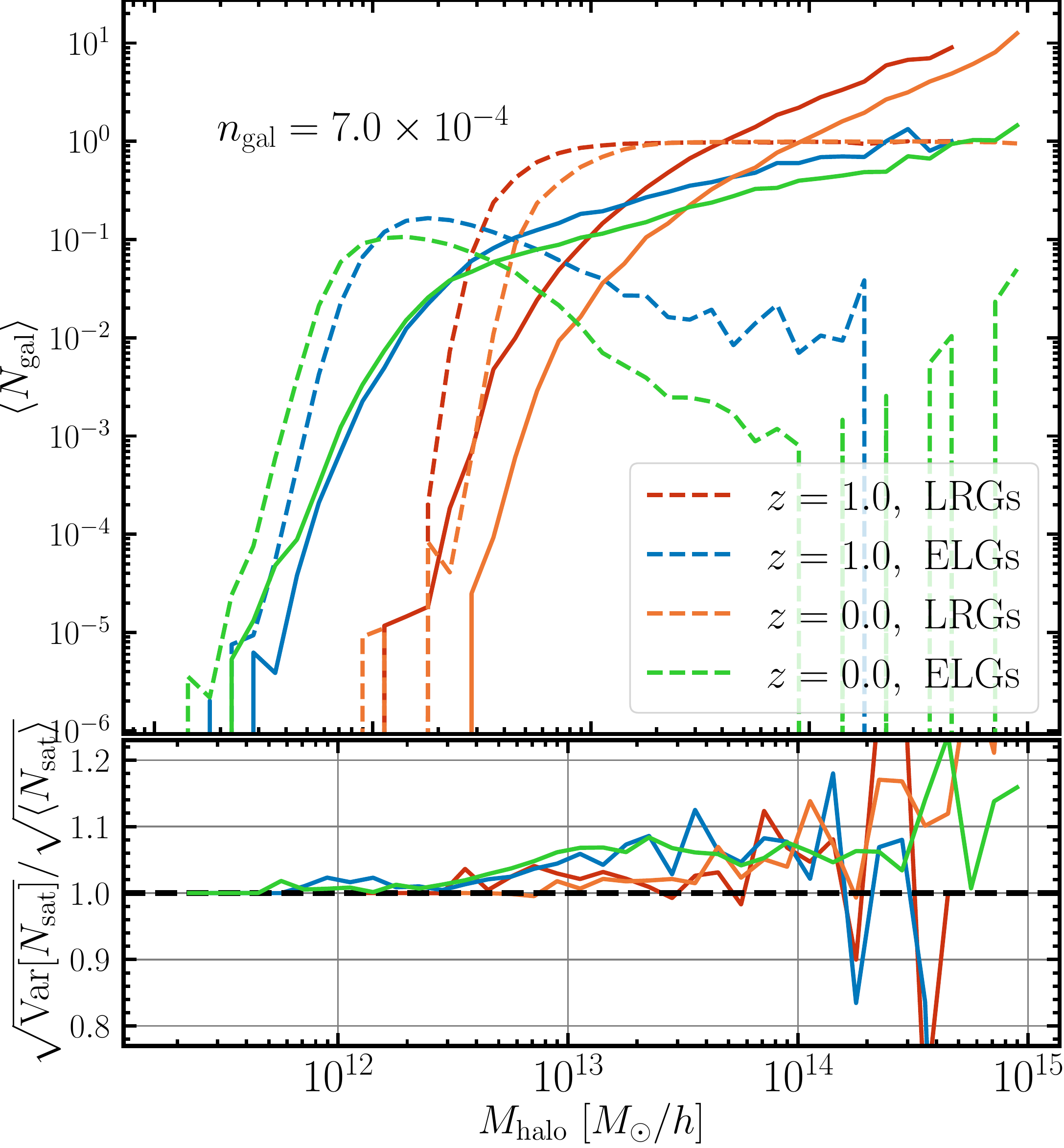} \\
\includegraphics[width=0.48\textwidth]{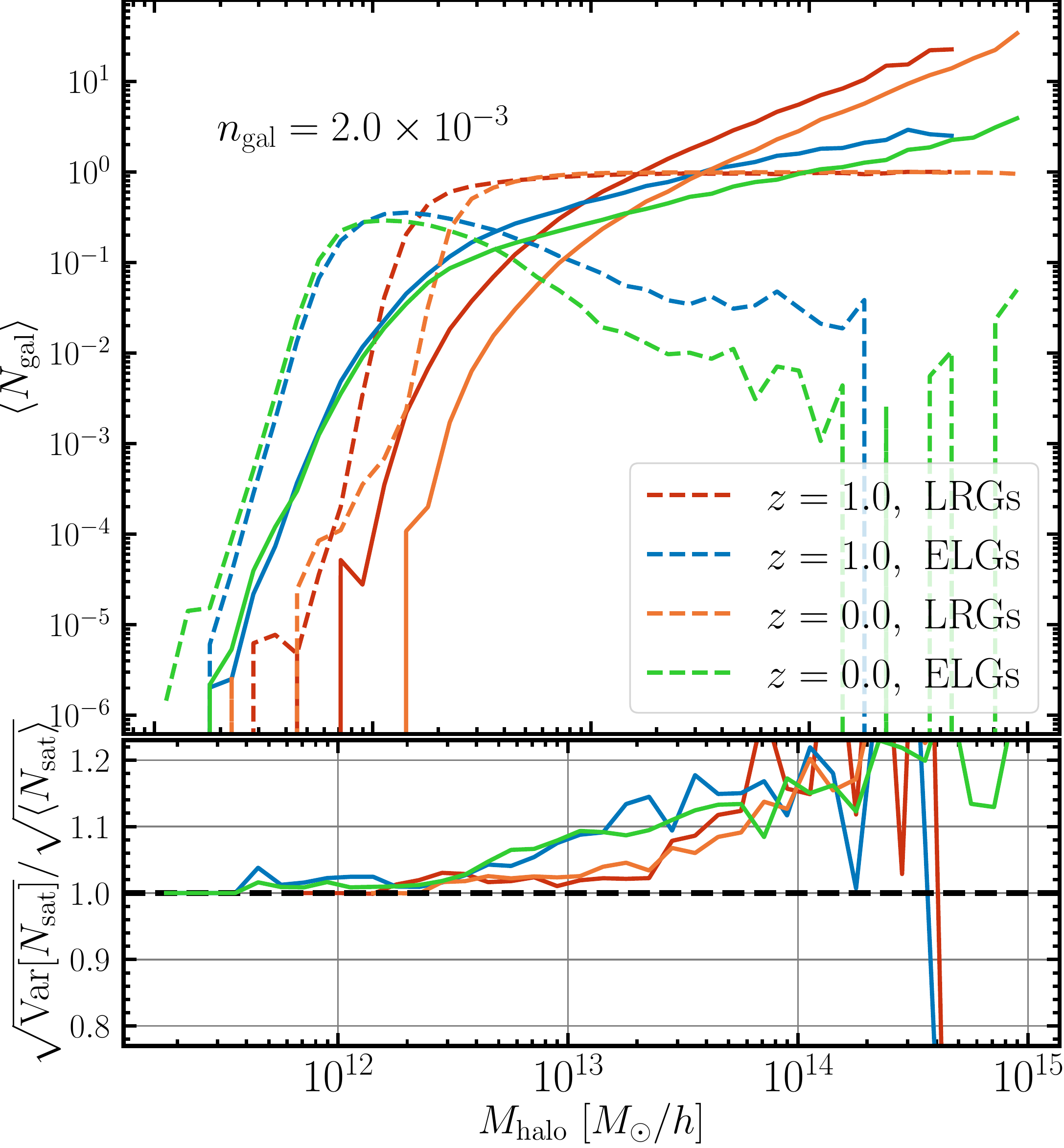}
\caption{Halo occupation distribution of the extracted LRGs and ELGs from the MillenniumTNG hydrodynamical simulation at redshifts $z = 0$ and $z=1$ at two different galaxy number densities, $n_{\rm gal} = 7.0 \times 10^{-4}$ (``low,'' \textit{top}) and $2.0 \times 10^{-3} \, (\hMpc)^{-3}$ (``high,'' \textit{bottom}). The top panel of each figure shows the mean occupation distribution, split into its central (dashed line) and satellite (solid line) contributions, while the bottom shows the ratio between the standard deviation of the satellite halo occupations and its square-root (i.e.~the Poisson prediction for the standard deviation). Strikingly, we find a clear departure from the Poisson prediction in the case of ELG satellites, which warrants careful modeling of the halo population. At the ``high'' number density, we see that the deviation from the Poisson prediction of both the $z = 0$ and $z=1$ ELG samples is roughly 10-15\%, which can have significant consequences for the one-halo term.}
\label{fig:hod}
\end{figure}

In Fig.~\ref{fig:hod}, we show the halo occupation distribution of the LRG and ELG samples we extract from MTNG at $z = 0$ and $z = 1$, at the ``low'' ($n_{\rm gal} = 7.0 \times 10^{-4}$) and ``high'' ($2.0 \times 10^{-3} \ [\hMpc]^{-3}$) galaxy number densities. Each of the panels shows the mean occupation distribution for the different samples, split into their central and satellite contributions. The lower panels give the ratio between the standard deviation of the satellite halo occupations and their square-root, which is the predicted standard deviation of the satellite halo occupations provided that they follow a Poisson distribution. This ratio allows us to explore deviations from the typically assumed Poisson halo occupation distribution for the satellites.

According to Fig.~\ref{fig:hod}, the shapes of the mean halo occupancy match well the qualitative form of the \citet{Zheng:2004id} and HMQ  \citep{2020MNRAS.497..581A} model predictions for the LRGs and ELGs, respectively. In addition, the behavior of each galaxy sample is similar for the two redshifts we consider. Interestingly, we find that as we go from high to low redshift, the masses of haloes containing an ELG central shift towards lower halo mass, whereas in the case of LRGs, the opposite is true. This is a reflection of the fact that haloes become on average more quenched and more massive as time progresses. Thus, by $z = 0$, ELG centrals living in more massive haloes have already been quenched, whereas those in less massive haloes are still undergoing vigorous star formation. As we will see in later sections, this has important implications for the clustering and bias of ELGs.

The most striking finding in Fig.~\ref{fig:hod} is the evident discrepancy from the Poisson prediction in the case of ELG satellites, which warrants careful modeling when populating haloes with galaxies. At the ``high'' number density, we see that the inconsistency with the Poisson prediction of both the $z = 0$ and $z=1$ ELG samples is roughly 10-15\%, which can have significant consequences for the one-halo term (see Section~\ref{sec:non_poiss}). The deviation for the ``low''-density ELG samples is less pronounced, at the 5-10\%-level, but still more noticeable than that of the LRGs. Nonetheless, the digression from Poisson variance for both LRGs and ELGs seems to increase with halo mass. 

While previous studies have similarly found a slight deviation in the case of red (LRG-like) and subhalo samples \citep[e.g.,][]{2017MNRAS.472..657J}, to our knowledge this is the first study that demonstrates that the deviation from the Poisson prediction is much more noticeable for ELGs (blue, star-forming galaxies). Work by \citet{2019MNRAS.490.3532J} also examined this question for both star-forming and quenched galaxies, but found no significant non-Poissonian signal. We attribute this to a difference in the galaxy selection procedure of star-forming galaxies (based on SFR rather than sSFR) and a different galaxy formation model (SAM rather than a hydro simulation).

\subsubsection{Pseudo-Poisson satellite occupation statistics}
\label{sec:non_poiss}

In Fig.~\ref{fig:hod}, we showed that the satellite occupation distribution appears to deviate from the traditionally assumed Poisson variance for some samples. In particular, we reported that the ELG samples, especially the ``high''-density ones, exhibit super-Poisson behavior in the halo mass range of $\sim 10^{13} \ \hMsun$ and above. In this Section, we discuss how the super-Poisson variance of ELG occupations affects the one-halo term and thus their two-point correlation. We then present a simple method to account for it by designing a new pseudo-Poisson distribution for modeling the satellite occupations. Finally, we compare the effect of employing the pseudo-Poisson distribution on the clustering of ELGs and LRGs.

In halo occupation modeling, one can write the one- and two-halo terms of the power spectrum (which is the analogous quantity to the two-point correlation function in Fourier space) in the following manner \citep{2010gfe..book.....M}:
\begin{equation}
    P^{1h}(k) = \int \mathrm{d}M \frac{\langle N (N-1) \rangle_M}{\bar n_g^2} n(M) \,|\tilde u_g(k|M)|^2,
\end{equation}
and
\begin{equation}
    P^{2h}(k) = P^{\rm lin}(k) \left[\int \mathrm{d}M \frac{\langle N \rangle_M}{\bar n_g} b(M) n(M) \tilde u_g(k|M)\right]^2 \, ,
\end{equation}
respectively, where $\langle N \rangle_M$ describes the average number of galaxies that reside in a halo of mass $M$, $\bar n_g$ is the average number density of those galaxies, $P^{\rm lin}(k)$ is the linear power spectrum, $n(M)$ is the halo mass function, $b(M)$ is the linear bias of haloes of mass $M$, and $\tilde u_g(k|M)$ is the Fourier transform of the normalized radial number density distribution of galaxies in haloes of mass $M$. Thus, the one-halo term of the galaxy auto-correlation function requires the second moment of the halo occupations, whereas the two-halo term depends solely on the mean. We can express the second moment in terms of the satellite and central occupation distributions as:
\begin{equation}
    \langle N (N-1) \rangle_M = \langle N_s (N_s - 1) \rangle_M + 2 \langle N_c \rangle_M \langle N_s \rangle_M \, ,
    \label{eq:NcNs}
\end{equation}
where we have assumed that the occupation statistics of satellites and centrals are independent, i.e. $\langle N_c N_s \rangle_M = \langle N_c \rangle_M \langle N_s \rangle_M$ (we will revisit this in Section \ref{sec:cond_prob}). The second moment of the satellite occupation distribution is:
\begin{equation}
    \langle N_s (N_s - 1) \rangle_M = \sum_{N_s = 0}^\infty N_s (N_s-1) P(N_s|M) \equiv \beta^2(M) \langle N_s \rangle^2 ,
\end{equation}
where we have introduced a new function $\beta$. If the occupation statistics of satellites follow Poisson statistics, i.e.
\begin{equation}
    P(N_s | M) = \frac{\lambda^{N_s} e^{-\lambda}}{N_s!}, \, \, \, {\rm where} \, \lambda = \langle N_s \rangle_M \, ,
    \label{eq:poiss}
\end{equation}
then $\beta(M) = 1$. If $\beta > 1$, the distribution is said to be super-Poisson (broader) and if smaller, it is sub-Poisson (narrower). We can now connect $\beta$ to the quantity measured in Fig.~\ref{fig:hod}, $\sqrt{{\rm Var}[N_s](M)}/\sqrt{\langle N_s \rangle_M}$  via:
\begin{equation}
    \beta^2(M) = \left(\frac{{\rm Var}[N_s]}{\langle N_s \rangle_M} - 1\right) \frac{1}{\langle N_s \rangle_M}+1 .
\end{equation}

Having shown that there is a noticeable deviation from Poisson statistics for some of our samples in Fig.~\ref{fig:hod}, which affects the one-halo term via the equations above,
we proceed to define an alternative distribution for the satellite occupations. We notice that if we scale both $\lambda$ and $N_s$ in Eq.~\eqref{eq:poiss} for the Poisson distribution by a free parameter $\alpha$, so that the distribution takes the form
\begin{equation}
    P(N_s | M) = \frac{(\alpha \lambda ) ^{\alpha N_s} e^{-\alpha \lambda}}{\Gamma({\alpha N_s + 1})}, \, \, \, {\rm where} \, \lambda = \langle N_s \rangle_M \, ,
\label{eq:pseudo}
\end{equation}
then the overall shape and the mean of the distribution ($\lambda = \langle N_s \rangle$) are retained, but the variance becomes $\lambda / \alpha$, where $\alpha > 1$ ($\alpha < 1$) corresponds to a sub-(super-)Poisson distribution. We have also swapped the factorial in the denominator with the more general Gamma function (note that $\Gamma (x+1) = x!$ for $x$ being a positive integer) because its argument, $\alpha N_s$, is no longer guaranteed to be an integer. 

Importantly, we can obtain an empirical estimate of $\alpha$ from the measured satellite occupation distributions as:
\begin{equation}
    \hat \alpha(M) = \frac{\langle N_s \rangle_M}{{\rm Var}[N_s]} ,
\end{equation}
which can be used when generating mock catalogues. Next, we consider how our choice of the pseudo-Poisson distribution of Eq.~\eqref{eq:pseudo} affects the one-halo term of galaxy clustering. For simplicity, and to reduce the number of free parameters in the model, we assume $\alpha$ to be constant over some mass range and equal to one outside that range.

\begin{figure}
\centering  
\includegraphics[width=0.48\textwidth]{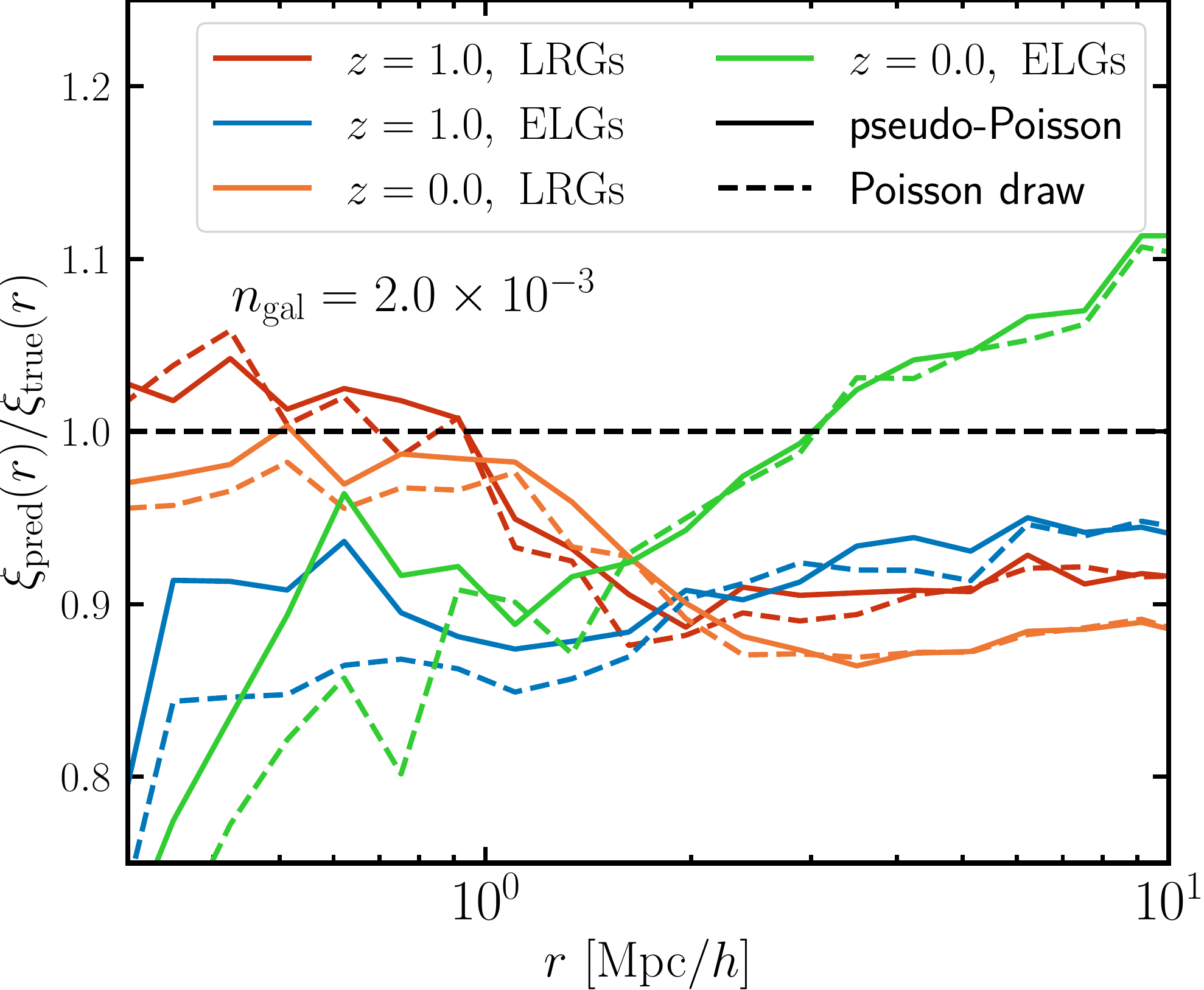}
\caption{Correlation function ratio between the predicted LRG and ELG samples at $z = 1$ and $z=0$, compared to  the ``true'' LRG and ELG samples extracted from MTNG. Results are shown for the ``high'' number density. The predicted samples are generated by adopting a mass-only HOD with mean halo occupations as a function of mass taken directly from the full-physics simulation. The solid lines indicate the result when the number of satellites in a halo were chosen by drawing from the pseudo-Poisson distribution of Eq.~\eqref{eq:pseudo}, while the dashed lines correspond to the traditionally used Poisson distribution. While in the case of the LRGs, switching to the pseudo-Poisson satellite distribution has a negligible effect on the clustering, in the case of ELGs, the clustering is increased by $\sim$10\% in the one-halo regime, improving the agreement with the ``true'' sample. The deviation at large distances is produced by assembly bias, which is studied in more detail in a companion paper \citep{Hadzhiyska2022b}.}
\label{fig:pseudo}
\end{figure}

In Fig.~\ref{fig:pseudo}, we show the ratio of the correlation function between the predicted LRG and ELG samples at $z = 1$, and the ``true'' LRG and ELG samples extracted from MTNG. We show the results only for the ``high'' number density, since Fig.~\ref{fig:hod} indicates that it is the ``high''-density sample that exhibits more noticeable pseudo-Poisson behavior. The predicted samples are generated by adopting a mass-only HOD (i.e., the most ``basic'' approach) with mean halo occupations as a function of mass taken directly from the full-physics simulation. The number of centrals in each halo is then drawn from a binomial distribution, whereas the number of satellites is either drawn from a Poisson distribution with mean taken from the HOD of satellites, or from a mixture between Poisson and pseudo-Poisson (Eq.~\ref{eq:pseudo}) distributions.

We obtain the free parameter $\alpha$ that appears in Eq.~\eqref{eq:pseudo} by calculating the minimum halo mass, $M_{{\rm min}}$, at which $\langle N_s \rangle/\rm{Var}[N_{\rm s}]$ first dips below 0.9\footnote{The threshold value of 0.9 is arbitrarily chosen, and its main purpose is to reduce the number of haloes for which we need to draw from the pseudo-Poisson distribution, which makes the occupation assignment more efficient.} and compute the average value of $\alpha$ between $M_{{\rm min}}$ and $M_{\rm max}$\footnote{$M_{\rm max}$ is the highest halo mass at which we apply the halo occupation model.
}. For the ELGs, the corresponding values of the two parameters are $\alpha \approx 0.8; \ M_{{\rm min}} \approx 9.0 \times 10^{12} \hMsun$, while for the LRGs, they are $\alpha \approx 0.80; \ M_{{\rm min}} \approx 5.7 \times 10^{13} \hMsun$. We note that we draw from the pseudo-Poisson distribution only for the haloes in the mass range $(M_{{\rm min}}, \ M_{\rm max})$, which decreases the computational cost of the halo assignment. Once the number of satellites and centrals has been determined, we distribute the satellites by following the steps in Section~\ref{sec:sat_model}; i.e. we draw randomly from the ``true'' satellite distribution at fixed mass. 

In Fig.~\ref{fig:pseudo}, we illustrate the effect of switching between Poisson and pseudo-Poisson satellite occupation distribution. In the case of the ELGs, their clustering is increased by $\sim$10\% in the one-halo regime ($r < 1 \ \hMpc$), improving the agreement with the ``true'' sample. In the case of the LRGs, the effect on the clustering is negligible. On larger scales, the clustering is unaffected, as expected. This finding suggests that halo models of ELGs may benefit from implementing pseudo-Poisson schemes for populating haloes with satellites, especially when pursuing subpercent accuracy on small scales. The deviation from one that we see on large scales is caused by the galaxy assembly bias effect; i.e. the dependence on additional halo properties apart from mass, which is neglected in the mass-only model used to generate the predicted samples \citep[this is discussed in more detail in our companion paper,][]{Hadzhiyska+2022b}. The remaining deficiency in the predicted ELG clustering is further discussed in Section~\ref{sec:sat_model}.

\subsubsection{Correlation of the central-satellite occupation statistics}
\label{sec:cond_prob}

Another common assumption of the halo occupation model is that the occupation numbers for centrals and satellites in a halo are independent of each other. This question has not been extensively studied in the literature \citep[see][for a general discussion]{2016MNRAS.460.2552H}, though the authors of \citet{2019MNRAS.490.3532J} investigated this effect in a semi-analytic model (SAM) for galaxy formation through accounting for the central-satellite occupation correlation, but they did not find a significant evidence for it. We note that both their galaxy formation model (SAM rather than hydro simulation) and the selection procedure of star forming galaxies (based on SFR rather than sSFR) differ from the present work. In this section, we challenge the assumption of independence of the central and satellite occupations in MTNG and propose a simple method for tying together the central and satellite probabilities. To our knowledge this is the first study of this issue in hydro simulations.

In this section, we concentrate only on the ELG samples, as the LRG selection procedure we adopt has a trivial central-satellite correlation. Specifically, it is obtained by making a cut in stellar mass; thus, in virtually all cases, the central galaxy has the highest stellar mass, and its presence is effectively a prerequisite for the ability of its parent halo to host satellites. We note that often some coupling of the central and satellite occupations in LRG samples is imposed by multiplying Eq.~\eqref{eq:nsat} by a Heaviside function, $\Theta(M_{\rm min})$, which ``bans'' the existence of satellites from lower-mass haloes that are incapable of even hosting a central. This modification is also often adopted when populating mock catalogues with other red and magnitude-limited galaxy samples. However, in the case of the ELGs, the central occupation distribution has a more complicated shape (see Fig.~\ref{fig:hod}), such that haloes are not guaranteed to have a central (in fact, the HOD peaks at $\sim$10\% before rapidly declining). This is the case because ELGs
are extremely sensitive to quenching, star formation, galactic cannibalism, tidal disruption, etc. Many of these processes are the direct or indirect result of central-satellite interactions, and therefore, make it more likely that a central-satellite correlation is induced in the occupation statistics.  The implication is that the properties of satellites (stellar mass, color, etc.) may be correlated with the properties of their centrals, which is also interesting to understand from an astrophysical point of view. The most significant effect this correlation would have on the large-scale galaxy statistics is a modification of the one-halo term (see Eq.~\ref{eq:NcNs}).

\begin{figure}
\centering  
\includegraphics[width=0.48\textwidth]{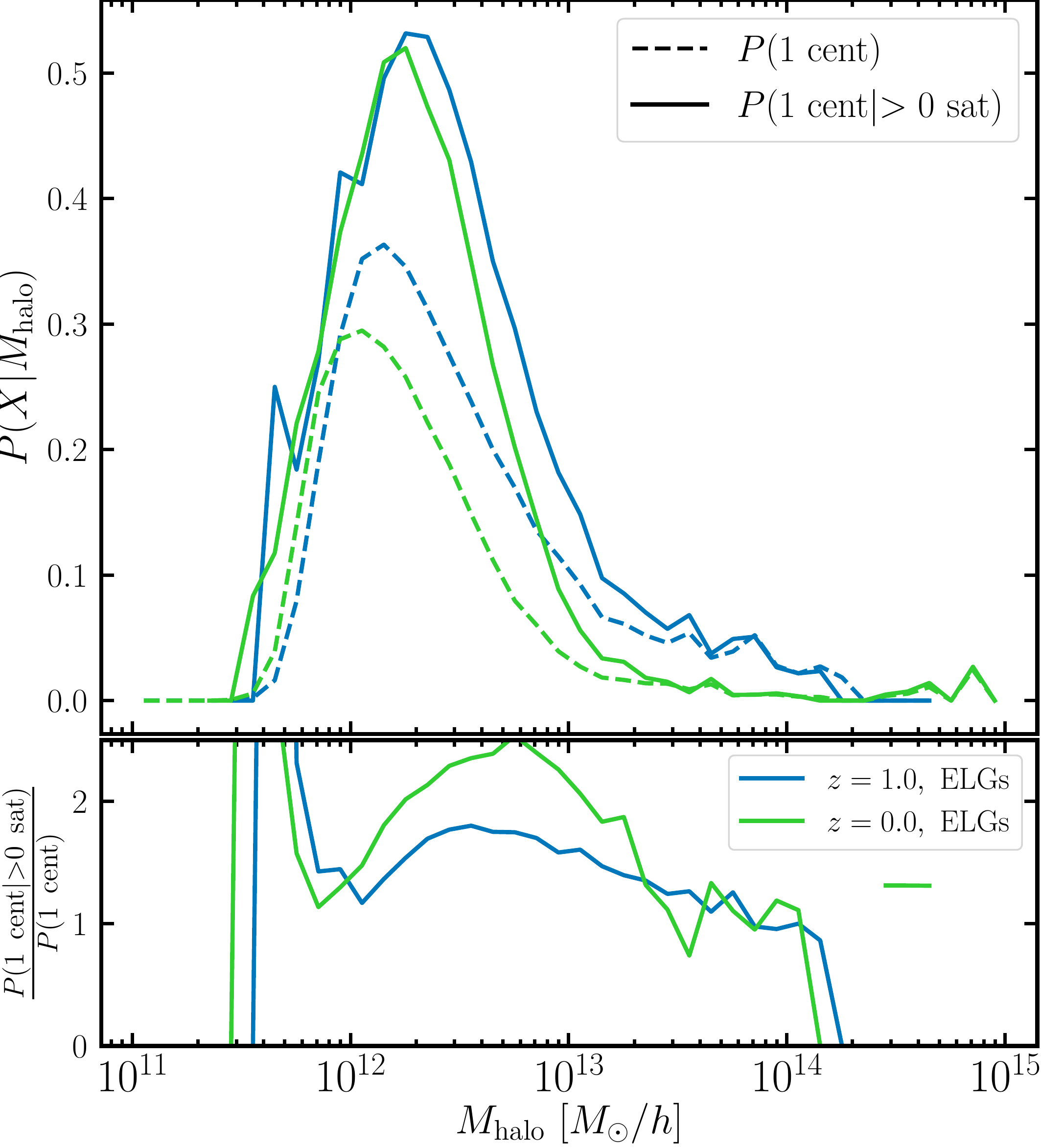}
\caption{Probability distribution of the ELG centrals at $z = 1$ and $z=0$ as a function of halo mass. The dashed curves indicate the probability that a halo of a given mass contains a central, $P(1 \ \rm cent)$ (matching the dashed curves in Fig.~\ref{fig:hod}). The solid lines correspond to the conditional probability, $P(1 \ {\rm cent}|>0 \ {\rm sat})$, that a halo contains a central given that it hosts one or more satellites. The lower panel shows the ratio of the solid to the dashed curves for each sample. Interestingly, we find that the probability that a halo has an ELG central given that it has at least one ELG satellite is about twice as large as the probability that it has a central with any satellite configuration over the mass range $10^{12}-10^{13} \, \hMsun$, which contains the majority of galaxies. The ratio peaks at halo mass $2 \times 10^{12}$ and $5 \times 10^{12} \, \hMsun$, and its maximum values are 1.8 and 2.5, for the $z = 1$ and $z=0$ samples, respectively.}
\label{fig:prob_cen}
\end{figure}

In Fig.~\ref{fig:prob_cen}, we show the probability distribution of the ELG centrals at $z = 1$ and $z=0$ as a function of halo mass. The curves indicating the probability that a halo of a given mass contains a central, $P(1 \ \rm cent)$, are equivalent to the dashed curves in the lower panel of Fig.~\ref{fig:hod}. The lines corresponding to the conditional probability, $P(1 \ {\rm cent}|>0 \ {\rm sat})$, signify the probability that a halo contains a central given that it hosts one or more satellites. We compute this quantity as the probability that a halo contains a central and at least one satellite, divided by the probability that it contains at least one satellite. Traditionally, the assumption is that the two probabilities are the same, i.e. $P(1 \ {\rm cent}|>0 \ {\rm sat}) = P(1 \ \rm cent)$. Interestingly, from this figure, we find that this is not the case. In fact, the probability that a halo has an ELG central given that it has at least one ELG satellite is about twice as large as the probability that it has a central with any satellite configuration, in the mass regime $10^{12}-10^{13} \ \hMsun$. The ratio peaks at halo masses $2 \times 10^{12}$ and $5 \times 10^{12} \ \hMsun$, with a maximum value of 1.8 and 2.5, for the $z = 1$ and $z=0$ samples, respectively.

We also notice that the conditioned curves, $P(1 \ {\rm cent}|>0 \ {\rm sat})$, are slightly offset from the total central occupation curves, $P(1 \ \rm cent)$. 
We denote their ratio by a new variable $\kappa(M)$, defined as:
\begin{equation}
    \kappa(M) \equiv \frac{P(1 \ {\rm cent}|>0 \ {\rm sat})}{P(1 \ \rm cent)} \, ,
\end{equation}
where the dependence on halo mass of the probabilities is implicitly assumed. We next explore the ramifications of this finding on the one-halo clustering, and provide some possible explanations.

To implement the conditional probability of ELG centrals into a mock creation model, we propose the following procedure. First, we use Eq.~\eqref{eq:NHMQ} and Eq.~\eqref{eq:nsat} to obtain the mean central and satellite occupations for each halo, given its mass. Next, we draw from a Poisson or pseudo-Poisson distribution (see Section~\ref{sec:non_poiss}) to determine the number of satellites that occupy each halo. We then split the haloes into two groups: haloes containing at least one ELG satellite, and haloes containing exactly zero satellites. To determine whether the haloes in \textit{the first group} contain a central or not, we flip a biased coin with probability $P(1 \ {\rm cent}|>0 \ {\rm sat}) = \kappa(M) P(1 \ \rm cent)$. For simplicity and to reduce the number of free parameters, we treat $\kappa$ as constant and set it to a fixed value, determined by averaging $\kappa(M)$ over the mass range of interest (though the calculations below hold regardless of whether $\kappa$ is a constant or a function of halo mass). From here onwards, we set $\kappa = 1.5$ and 1.8 for the $z = 1$ and $z = 0$ ELG samples, respectively.

For \textit{the remaining haloes}; i.e.~those containing zero satellites, we need to express the conditional probability of having a central given no satellites, $P(1 \ {\rm cent}|0 \ {\rm sat})$, through the parameters in the HOD model: the central and satellite probabilities, $P(1 \ \rm cent)$ and $P(>0 \ {\rm sat})$, and the conditional probability $P(1 \ {\rm cent}|>0 \ {\rm sat}) = \kappa P(1 \ \rm cent)$. We arrive at the following expression:
\begin{equation}
    P(1 \ {\rm cent}|0 \ {\rm sat}) = \left[\frac{P(1 \ {\rm cent})}{P(>0 \ {\rm sat})} - P(1 \ {\rm cent}|>0 \ {\rm sat})\right] \frac{P(>0 \ {\rm sat})}{1-P(>0 \ {\rm sat})} \, ,
\end{equation}
which upon substituting $\kappa$ can be reworked into:
\begin{equation}
    P(1 \ {\rm cent}|0 \ {\rm sat}) = P(1 \ {\rm cent}) \left[1 + \frac{1-\kappa}{{P(>0 \ {\rm sat})}^{-1}-1} \right] \, .
\end{equation}
All probabilities are conditioned on the halo mass, but the dependence is not denoted for the sake of brevity. Having determined the occupation numbers of both the centrals and the satellites in each halo, we can proceed with the creation of mocks. We first need to adopt a one-halo population scheme. For this study, we choose the phenomenological approach for the ELGs in Section~\ref{sec:sat_prop}, which accounts for the observed anisotropy of the satellite distribution.

\begin{figure}
\centering  
\includegraphics[width=0.48\textwidth]{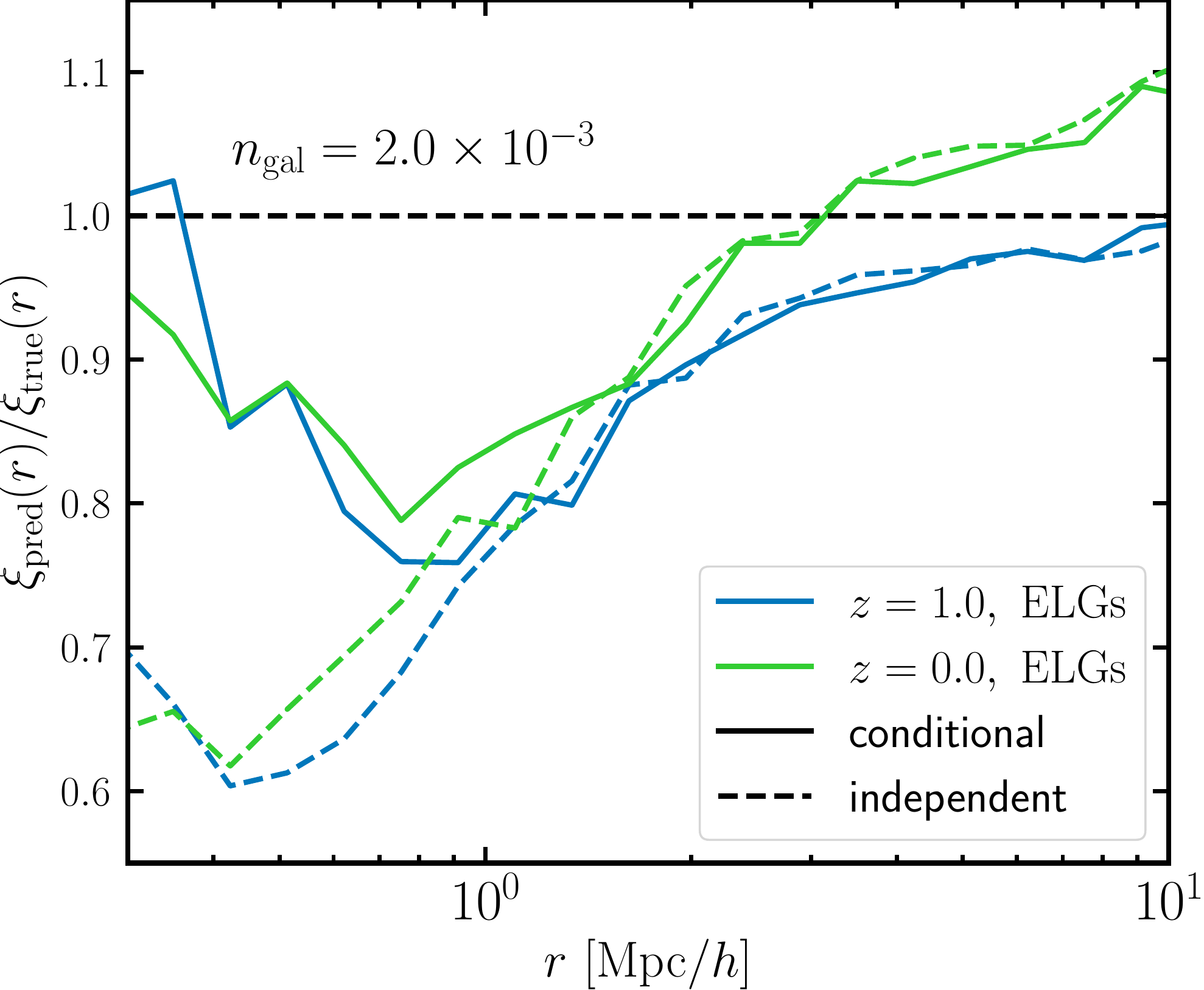}
\caption{Correlation function ratio between the predicted and ``true'' high-density ELG samples at $z = 1$ and $z=0$. The predicted samples are generated by either assuming that the satellite and central occupation distributions are independent (dashed lines) or by adopting the conditional probability approach developed in Section~\ref{sec:cond_prob} (solid lines). The satellite occupations are obtained via Poisson draws, and the satellite profile distribution follows the procedure in Section~\ref{sec:sat_prop}. For the mean occupations, we adopt the mass-only HOD scenario for simplicity, which affects the two-halo regime and is easy to disentangle from the satellite-central occupation correlation. We see that both samples assuming a central-satellite correlation show substantial improvement over the samples assuming independent central and satellite probabilities. While the large-scale behaviour is unchanged ($r > 1 \ \hMpc$), as expected, the one-halo term of the central-satellite correlated sample appears to match reasonably well the ``true'' one-halo term at $r \approx 0.1 \ \hMpc$. However, we observe that it starts to deviate near the edges of the halo, approaching the results for the independently drawn samples near $r \approx 0.7 \ \hMpc$. We conjecture that this is related to the effect of anisotropic satellite distributions of the ELGs discussed in Section~\ref{sec:ang_dist}.}
\label{fig:cond_prob}
\end{figure}

In Fig.~\ref{fig:cond_prob}, we show that there is a noticeable improvement to the one-halo term when we adopt the conditional probability procedure outlined above, compared with the traditional approach which assumes independence of the central and satellite occupation distributions. In particular, the ratio of the correlation functions for the conditional probability model is close to one; i.e. perfect agreement, around $r \approx 0.1 \ \hMpc$, but begins to deviate near the edges of the halo, converging with the prediction of the standard model near $r \approx 0.7 \ \hMpc$. We conjecture that this is related to the effect of the anisotropic satellite distribution of the ELGs discussed in Section~\ref{sec:ang_dist}. 

A plausible interpretation of the correlation between central and satellite occupations is provided by the cooperative galaxy formation hypothesis. A more commonly used term in the literature in recent years is ``galaxy conformity,'' first detected in SDSS data \citep{2006MNRAS.366....2W} and later investigated theoretically in \citet{2013MNRAS.430.1447K} and
\citet{2015MNRAS.454.1840K} \citep[see also][]{2015MNRAS.452.1958H, 2016MNRAS.461.2135H, 2017MNRAS.470.1298P, 2018MNRAS.480.2031C, 2020MNRAS.492.2722O, 2022MNRAS.511.1789Z}. According to the cooperative galaxy formation model, first proposed in \citet{1993ApJ...405..403B}, cooperative effects arise through radiative and hydrodynamical processes during protogalactic evolution and contribute beyond the traditional models of galaxy formation, which predict that the correlatedness in the luminosity of nearby galaxies due to the large-scale correlation between their density peaks (i.e., ``proto-supercluster'' and ``proto-void'' scales). According to the galaxy conformity conjecture, there are additional non-gravitational and non-local effects, which make it more likely that we find brighter galaxies in the presence of other bright galaxies nearby. An example is the collapse of proto-structures into filaments and pancakes, which could result in an increase in the pressure and density of the intergalactic medium and an enhancement of the galaxy formation processes.

Another possible mechanism for achieving coherent galaxy formation is through large-scale star formation, which would be accompanied by massive superwinds. These superwinds might then propagate towards neighbouring protogalactic clouds and stimulate further galaxy formation. This supposition is related to the idea of ``explosive galaxy formation'', through which galaxies form and produce stars by triggering each other's activity in quick succession.

After roughly $\sim$1 Gyr, we expect ELGs to become quenched and turn into red galaxies. By that time, they have also drifted apart from each other, which might explain why the ELGs exhibit stronger spatial correlation compared with the LRG satellites. We explore this in detail in Section~\ref{sec:ang_dist}, where we test for anisotropy in the satellite distribution. Cooperative galaxy formation may also manifest itself in the form of environmental assembly bias, which we address in our companion paper \citep{Hadzhiyska+2022b}. A more detailed test of this hypothesis is in any case warranted to elucidate the physical mechanism of the coherent central-satellite formation process we observe in MTNG. This is however left for future work, as the aim of the current study is to merely identify problematic assumptions of the halo occupation model and propose viable solutions that can be straightforwardly implemented in current cosmological analyses.

\subsection{Satellite distribution model}
\label{sec:sat_model}

One of the key steps in creating mock galaxy catalogues is to decide how the satellite galaxies are distributed spatially within their host haloes. This is vital for recovering the small-scale clustering on scales of $r \lesssim 3 \ \hMpc$, corresponding to the one-halo regime and the transition between the one- and two-halo terms. Typically, satellite positions and velocities are assigned via one of three schemes: 
\begin{enumerate}
    \item based on assuming that satellites trace the dark-matter profile of the halo, one can fit a \citet*[][NFW]{Navarro1996} profile to the haloes in the simulation and draw randomly from it to populate each halo with satellites (angular positions being chosen randomly). The velocity of the satellites is typically simply drawn from a Gaussian using the halo's velocity dispersion. This method is computationally inexpensive, as one can obtain analytical predictions for the dark-matter profiles without accessing the particle data, but it gives an inaccurate correlation with the matter field of the simulation.
    \item a slight modification to the above is to place satellites on a randomly-selected dark matter particle. This ameliorates the issue with the lack of a tight correlation between galaxies and matter, but one here needs to access the particle data for each halo. 
    \item picking a random subhalo within the halo (or doing abundance matching using some subhalo property). A problem with this method is that the subhaloes in a hydrodynamic simulation behave markedly differently from those in an $N$-body simulation, since tidal disruption and stripping are modified due to a modified halo structure, affecting the phase-space distribution and survival of satellites \citep[e.g.][]{Donghia2010,2014Natur.509..177V,2021MNRAS.501.1603H}. In addition, this method requires that we have identified subhaloes, which can be expensive for large simulations and depends both on the resolution of the simulation and the subhalo finder.
\end{enumerate}

In our model, we take advantage of having the knowledge of the ``true'' positions of the galaxies in MTNG when baryonic effects are fully accounted for. Thus, instead of drawing the satellite positions randomly from an NFW profile or from the particle/subhalo sample, we can draw from the ``true'' satellite distribution seen in the hydrodynamic simulation at fixed halo mass.

In detail, we can proceed in different ways. We can find the closest subhalo/particle to the drawn satellite radius, or we can assume that the satellites are isotropically distributed around the halo and assign a ``ghost'' galaxy. The advantage of using the nearest subhaloes/particles is that we trace the substructure inside the halo better, and we do not need to assume that its shape is spherical. However, an issue we discovered is that we cannot always find a subhalo/particle even within $100\,h^{-1} {\rm kpc}$ of the drawn radius, as some FoF groups exhibit percolation issues and their satellites lie outside the virial radius (see Fig.~\ref{fig:dist_prof}). Since this paper considers large-scale statistics of only the galaxies, we resort to the ``ghost'' galaxy method, which also gives us the freedom of switching on and off various features of our model and testing their effects on the galaxy summary statistics. 

\subsubsection{Radial profile}
\label{sec:rad_prof}

\begin{figure}
\centering  
\includegraphics[width=0.48\textwidth]{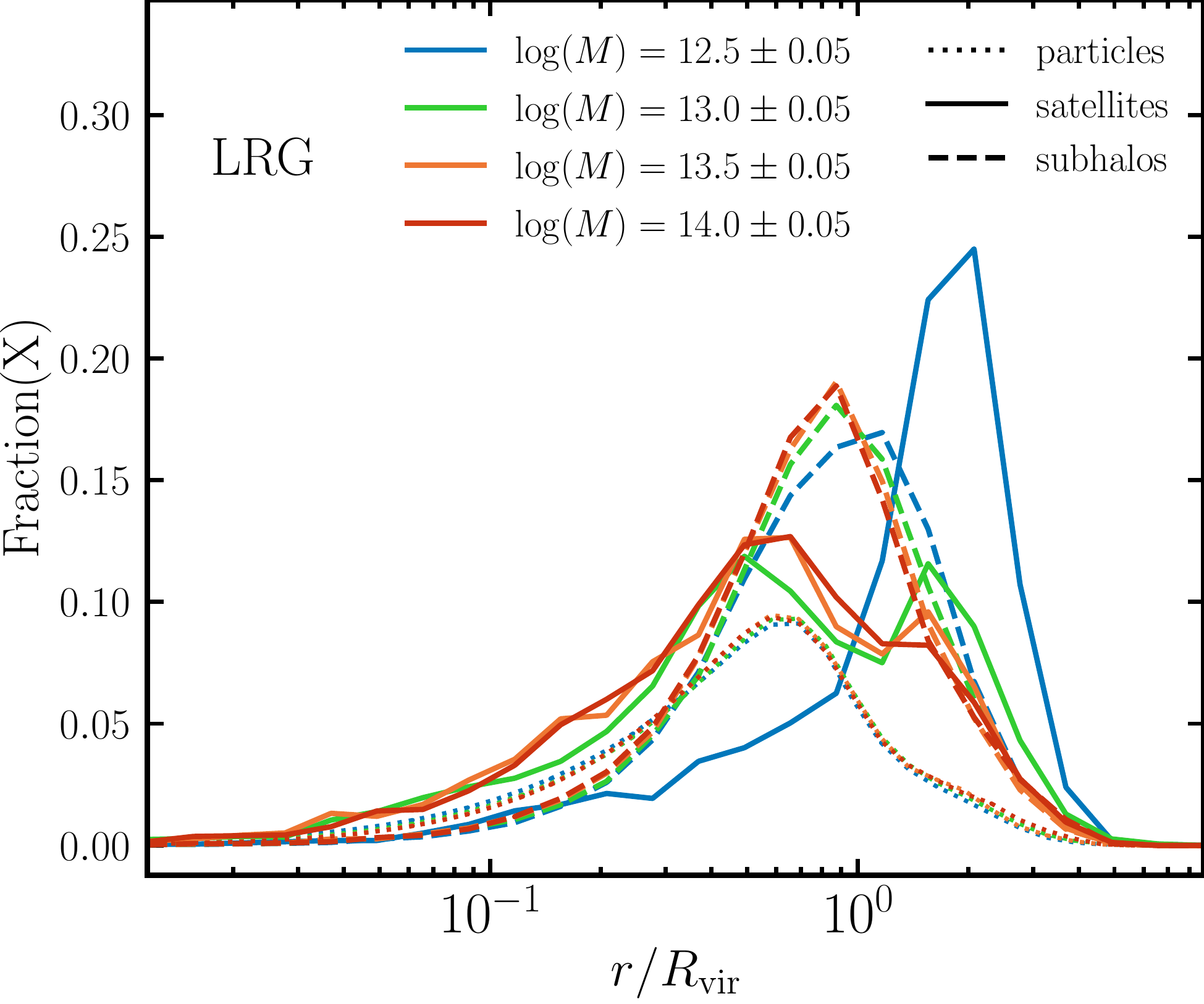}
\includegraphics[width=0.48\textwidth]{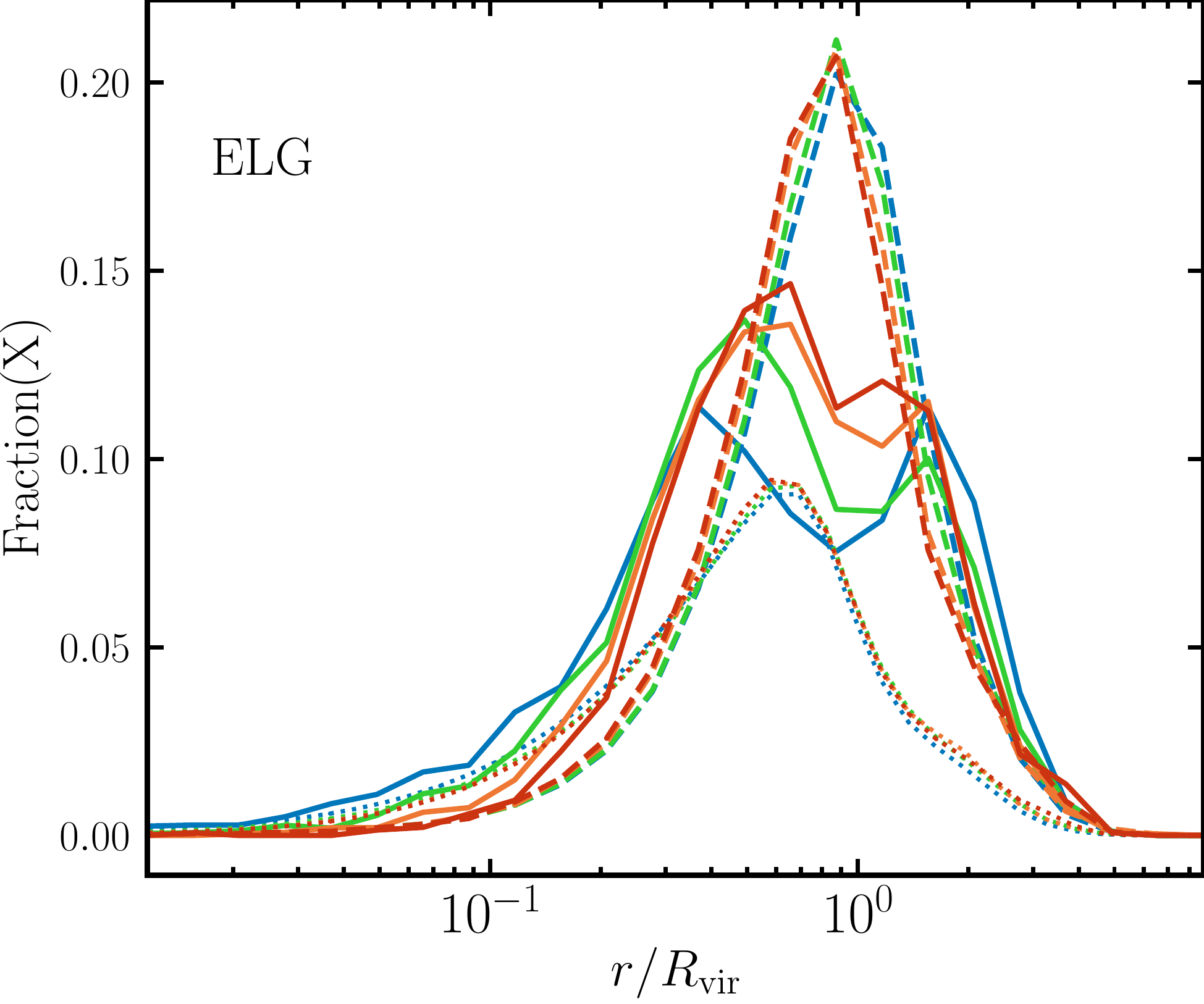}
\caption{Radial profiles of the galaxies (solid lines, ``high'' number density), subhaloes (dashed lines), and dark matter particles (dotted lines), at $z = 1$ for the full-physics MTNG run. We show LRGs (top panel) and ELGs (bottom panel) in four mass bins: [12.5, 13.0, 13.5, 14.0] $\pm$ 0.05 in units of $\log(M \ [\hMsun])$. The satellite profiles exhibit a noticeable bimodality; i.e.~a large number of satellites lies outside the virial radius and contributes to a second bump at $r/R_{\rm vir} > 1$. These are the result of FoF percolation; i.e.~two or more dark-matter clumps are strung together by feeble particle connections between halos. The effect is most pronounced for LRGs in the lowest mass bin, and almost disappears in the highest mass bin. The second bump for ELGs appears to be less halo-mass dependent. The particle profiles also exhibit bimodality, albeit of much smaller and less mass-dependent amplitudes, suggesting that FoF haloes are equally likely to percolate regardless of their mass. The subhalo profiles are slightly offset from the particle profiles, peaking at $r/R_{\rm vir} \approx 0.9$ rather than $r/R_{\rm vir} \approx 0.6$. The radial satellite distribution is, therefore, better traced by the particles, although they still underestimate the satellite population outside the virial radius.}
\label{fig:dist_prof}
\end{figure}

Next, we discuss the radial and velocity profiles of the ``true'' galaxies, dark matter particles, and subhaloes in the full-physics MTNG simulation, and summarize the proposed satellite distribution model. We first test the question of whether the subhaloes and particles trace the radial profiles of the satellite galaxies. In Fig.~\ref{fig:dist_prof}, we show the radial profiles of satellites, subhaloes, and dark matter particles at $z = 1$ for the full-physics MTNG run split into four mass bins: [12.5, 13.0, 13.5, 14.0] $\pm$ 0.05 in units of $\log(M \ [\hMsun])$. The satellites come from the LRG and ELG samples at the ``high'' number density. The particle and subhalo profiles are calculated for all haloes in a given mass bin regardless of whether or not they host a satellite galaxy. We select subhaloes with total mass above $\log (M) = 10$ for the ELGs and $\log(M) = 10.5$ (in units of $\hMsun$) for the LRGs, corresponding to the minimum total subhalo mass of each tracer.

The most striking feature of the satellite profiles is that they exhibit a bimodal behavior; i.e.~a large number of the satellites lie outside the virial radius and contribute to a second bump in the satellite profiles. Visualizing the haloes hosting these satellites\footnote{At \href{https://drive.google.com/drive/folders/1e4z1Psv_OmSJ_2KkYrpENLS2WqV_Zfvi?usp=sharing}{this link}, we have compiled two-dimensional visualizations of haloes of mass $10^{13}\, \hMsun$ containing satellites outside the virial radius.}, we see that they are often the result of FoF percolation; i.e., two or more dark-matter clumps are strung together by the FoF algorithm. This effect is the most pronounced for LRG satellites in the lowest mass bins and is much more diminished in the highest mass bin. The most plausible explanation is that haloes of mass $\log(M) = 12.5$ are very unlikely to host LRG satellites, as they are rarely massive enough to support a satellite (see Fig.~\ref{fig:hod}), and a satellite would either be tidally destroyed as it gets near the central, or merged into it (which would substantially increase the total halo mass and bump the halo to a higher mass bin). Therefore, the cases we are most likely to observe in that mass range involve two haloes in the process of merging, but which are already identified by the FoF algorithm as a single halo. Note that the virial mass of such a FoF group is much smaller than the FoF mass itself, as the virial radius cuts through the middle of these groups, effectively deblending them. The ELG satellite profiles, on the other hand, appear to be less halo-mass dependent, although they also display bimodal behavior.

While ELG-hosting subhaloes are not as massive as the LRG-hosting ones (compare $\log M > 10$ and $\log M > 10.5$), they still occupy relatively massive host haloes and appear to prefer the outskirts of the halo, which at a fixed virial mass, is more likely to occur in the case of percolating FoF groups, as their virial mass is then low compared with the total FoF mass. We conjecture that this preference arises because ELGs are quenched as they enter the higher density inner regions of the halo, so we are more likely to find them further away from the halo centre (we see further evidence for that in Section~\ref{sec:rad_vel}).

We can also see a reflection of the bimodality effect in the dark-matter particle profiles, which exhibit a bump outside the virial radius (albeit of much smaller amplitude). Interestingly, compared with the satellites, we barely find any mass-dependence in the particle profiles, suggesting that FoF haloes are equally likely to percolate regardless of their mass. The subhalo profiles, in addition to being insensitive to mass, do not show a bump outside their virial radius and are, in fact, slightly offset from the particle profiles, peaking at $r/R_{\rm vir} \approx 1$ rather than $r/R_{\rm vir} \approx 0.6$. It is evident from the figure that the radial satellite distribution is better traced by the particles, although if we place satellites on random particles, we would underestimate the satellite population outside the virial radius. We stress again that this is a feature of using the FoF finder and is likely to go away or be modified when either using just the gravitationally bound subhalos determined by {\small SUBFIND-HBT}, or non-FoF-based group finders, such as {\small ROCKSTAR} \citep{rockstar} or {\small CompaSO} \citep{2022MNRAS.509..501H}. We leave this question to a future study.

\subsubsection{Velocity dispersion profile}
\label{sec:vel_disp}

\begin{figure}
\centering  
\includegraphics[width=0.48\textwidth]{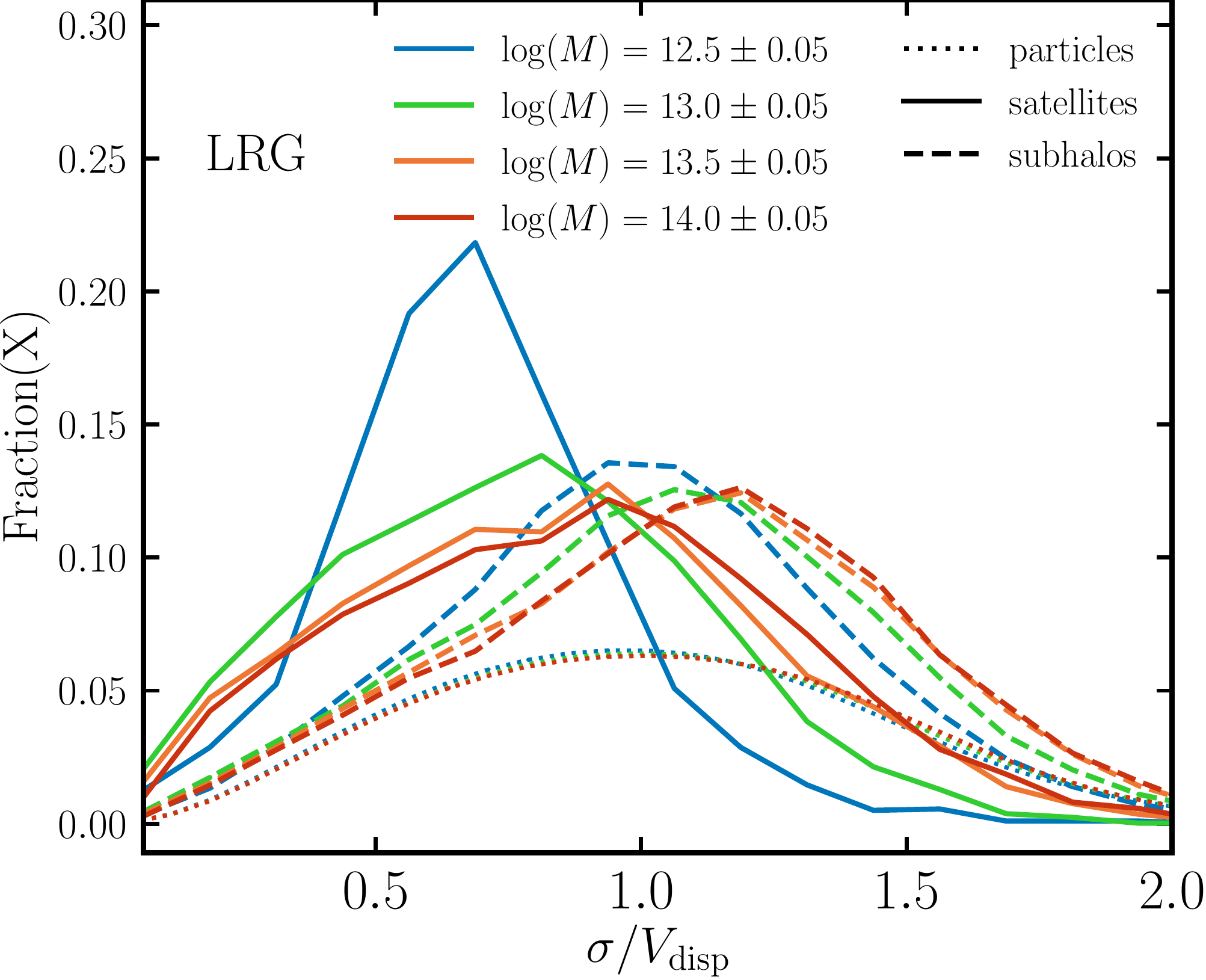}
\includegraphics[width=0.48\textwidth]{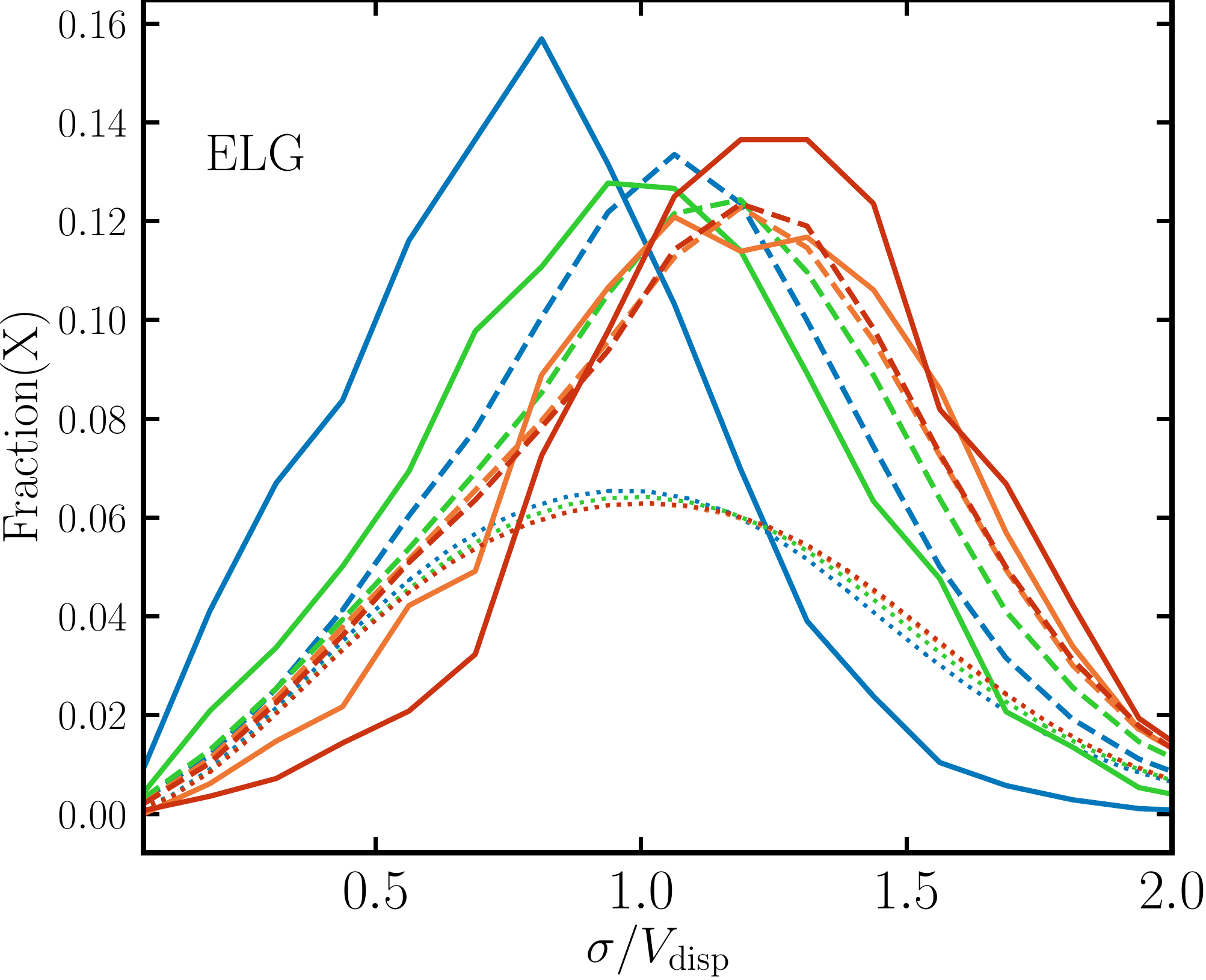}
\caption{Probability distribution of the velocity dispersion of galaxies (solid lines, ``high'' number density), subhaloes (dashed lines), and dark matter particles (dotted lines) at $z = 1$ for the full-physics MTNG run. We give results for the LRGs (top panel) and ELGs (bottom panel) in four mass bins: [12.5, 13.0, 13.5, 14.0] $\pm$ 0.05 in units of $\log(M \ [\hMsun])$. Both the satellites and the subhaloes exhibit a strong mass dependence, with the mean velocity dispersion shifting to larger values with mass. The ratio $\sigma/V_{\rm disp}$ is lowest for the LRGs in the lowest mass bin, which has important implications for mock construction, as it suggests that drawing from a Gaussian of width $V_{\rm disp}$ would make the galaxy velocities of LRGs hotter than they should be. On the other hand, we see that the ratio $\sigma/V_{\rm disp}$ exceeds the one for the ELGs in the highest mass bins, which could also lead to complications in mock creation. Similarly to our observation in Fig.~\ref{fig:dist_prof}, we find that in the highest mass bins the particles are better tracers of the satellite velocities than the subhaloes.}
\label{fig:disp_prof}
\end{figure}

In halo occupation modelling, one typically assumes that the velocities of satellites are Gaussian distributed with the velocity dispersion of the parent halo. We test this assumption by studying the histograms of the velocity dispersions of satellites, subhaloes and particles. Fig.~\ref{fig:disp_prof} gives the velocity dispersion histograms of galaxies, subhaloes, and dark matter particles at redshift $z = 1$ for the full-physics MTNG run, shown for the ``high'' density LRG and ELG samples in four mass bins: [12.5, 13.0, 13.5, 14.0] $\pm$ 0.05 in units of $\log(M \ [\hMsun])$. The curves are normalized to sum to unity, and the velocities of the satellites, subhaloes, and particles are normalized by the velocity dispersion of the halo they occupy. 

Overall, we see that the LRG and ELG satellites as well as the subhaloes exhibit a strong mass dependence, with the mean velocity dispersion shifting to larger values with increasing halo mass. We note that this shift is more noticeable for ELGs, which could constitute a challenge to reproduce the correct clustering in redshift space for simple population models that assign velocity to galaxies by drawing around a mean centered on the halo velocity dispersion. This is a manifestation of the same effect we see in Fig.~\ref{fig:dist_prof}, where for higher mass bins, the mean of the radial profiles shifts towards smaller radii, where the gravitational forces are stronger and the velocity dispersion is larger \citep[see e.g.,][for an early measurement of the dispersion profile of the Milky Way]{2010AJ....139...59B}. The ratio $\sigma/V_{\rm disp}$ is lowest for the LRG satellites in the lowest mass bin. Comparing this with the corresponding curve in Fig.~\ref{fig:dist_prof}, we conclude that this is caused by the population of satellites lingering right outside the halo virial radius.  This effect has important implications for mock construction. It suggests that according to MTNG, drawing from a Gaussian of width $V_{\rm disp}$ would make the galaxy velocities of LRGs hotter than they should be, which would artificially increase the power in the quadrupole moment of the correlation function. On the other hand, we see that the ratio $\sigma/V_{\rm disp}$ becomes larger than one for the ELGs in the largest two mass bins, which has the opposite effect of making the mock galaxies colder. This finding needs to be reexamined when adopting an alternative halo finder, as part of the effect is likely a consequence of FoF percolation.

Similarly to our observation in Fig.~\ref{fig:dist_prof}, we find that for the highest mass bins the particles are better tracers of the satellite velocities than the subhaloes. The mean of the subhalo velocity dispersion skews towards large $\sigma/V_{\rm disp}$ ratios, reflecting the migration of subhaloes towards the halo centre with increasing mass (Fig.~\ref{fig:dist_prof}). Recall that the satellite and subhalo velocities are calculated as the mean of all particles in the satellite-hosting subhalo. We find the least amount of mass dependence in the particle distributions, although there is a slight tendency towards a larger velocity dispersion with increasing mass, as with the other two tracers. This might be due to the fact that we use only the dark matter particles to compute the particle velocity dispersion rather than all particles (i.e.~with baryons). The effect of including the baryons, which tend to condense near the centers of halos and have a low dispersion velocity, is most prominent for haloes in the lowest mass bin, as the stellar-to-halo mass ratio peaks at $M \sim 10^{12} \, \hMpc$.

\subsubsection{Radial velocity profile}
\label{sec:rad_vel}

\begin{figure}
\centering  
\includegraphics[width=0.48\textwidth]{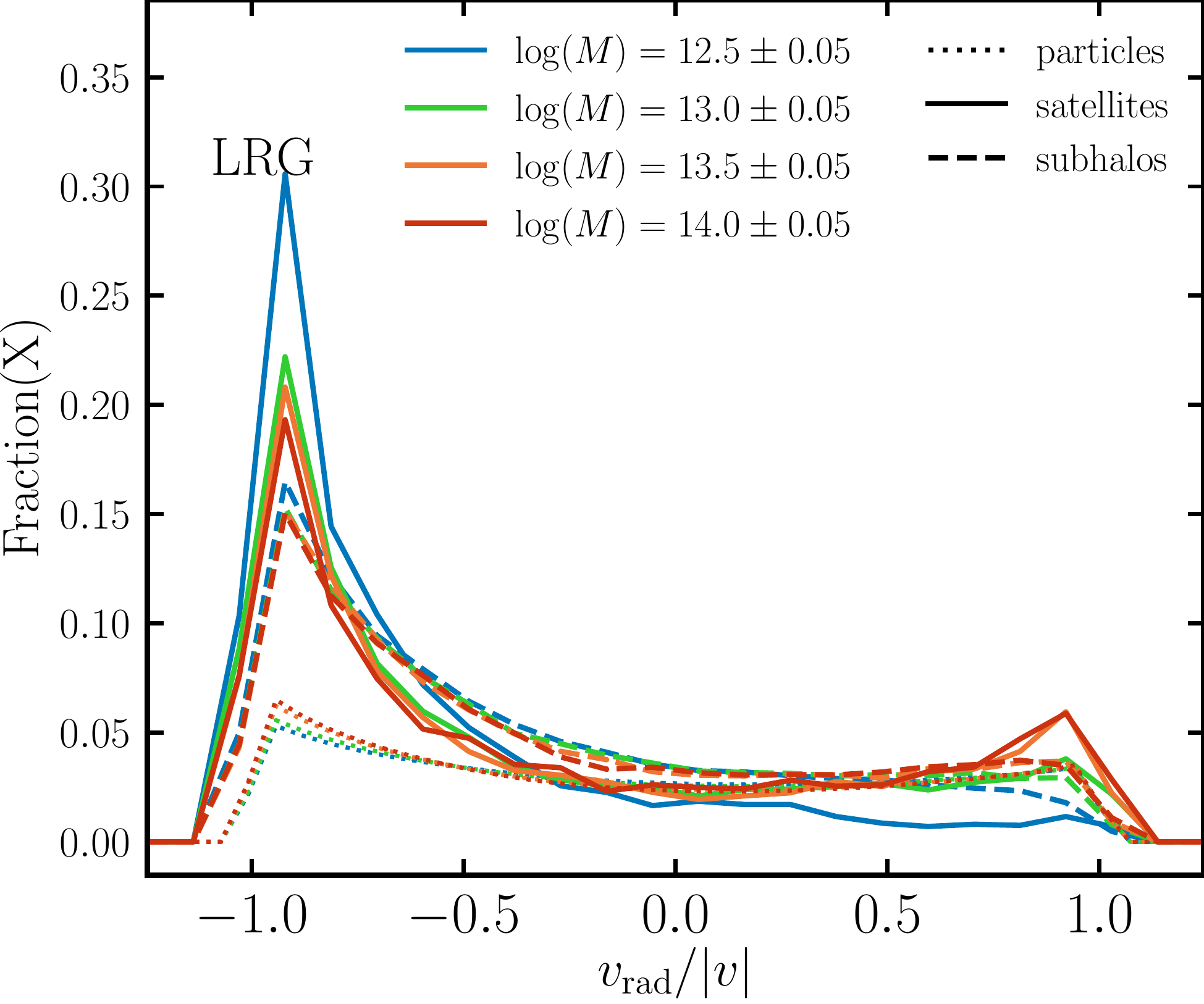}
\includegraphics[width=0.48\textwidth]{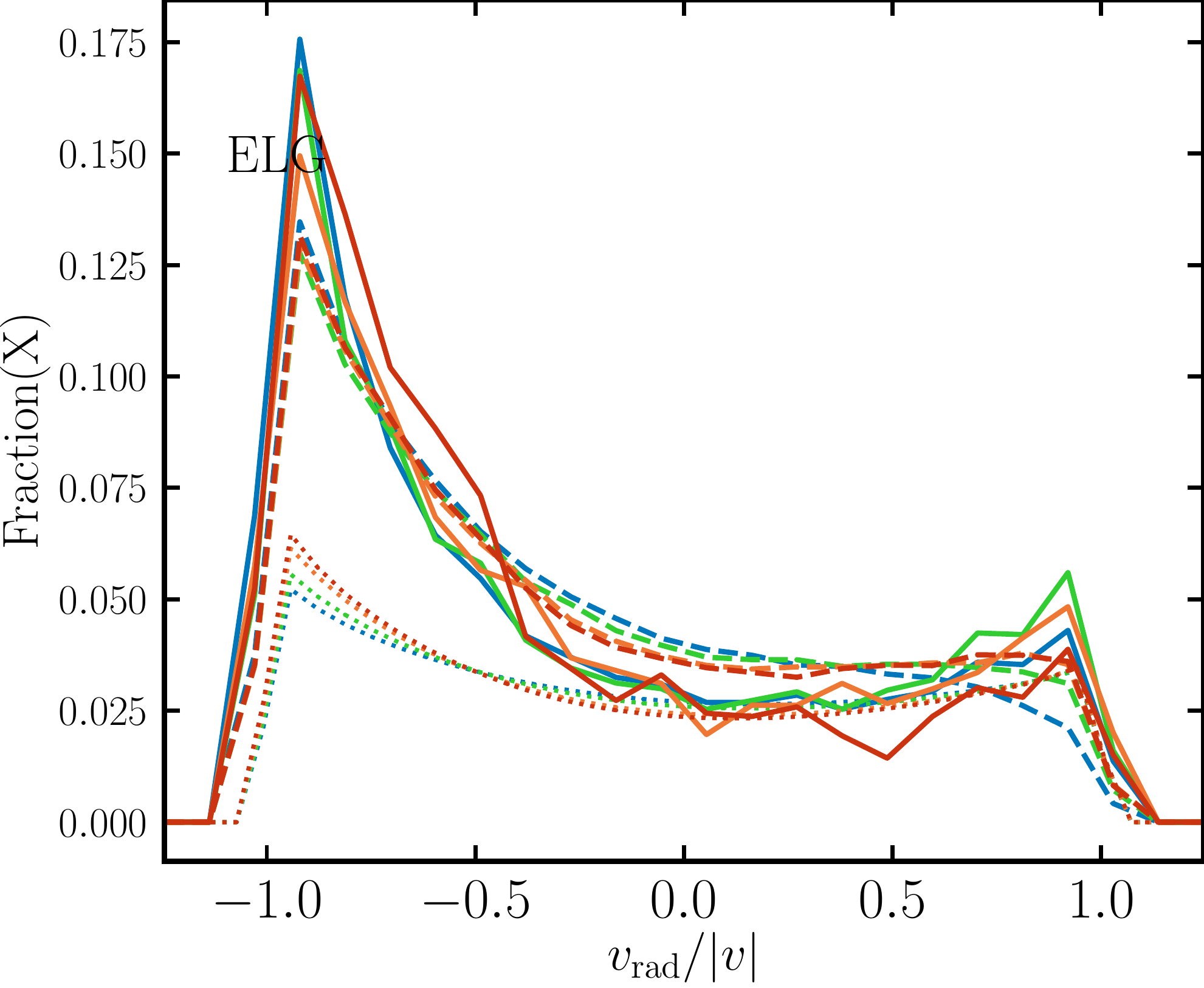}
\caption{Normalized radial velocity distribution of the galaxies (solid lines, ``high'' number density), subhaloes (dashed lines), and dark matter particles (dotted lines) at $z = 1$ for the full-physics MTNG run. We plot results for the LRGs (top panel) and ELGs (bottom panel) in four mass bins: [12.5, 13.0, 13.5, 14.0] $\pm$ 0.05 in units of $\log(M \ [\hMsun])$. All three types of objects exhibit a slight preference for asymmetric radial motion; i.e.~objects are more likely to be falling in than moving out, with the asymmetry being the most extreme for satellites and subhaloes. The LRG asymmetry decreases dramatically with mass: the smallest mass bin is the most extreme, with the majority of satellites being on inward radial orbits (presumably before merging onto the central). On the other hand, the highest mass bin of the ELG satellites exhibits the highest asymmetry. We conjecture this is the result of quenching, which is strongest in the highest mass bins. The particle distributions are the most symmetric, as particles cannot be destroyed, and any lingering asymmetry is the result of halo dynamics and resolution effects.}
\label{fig:vrad_prof}
\end{figure}

Typically, mock creation algorithms assign random velocity directions to the satellite velocities. Here, we test whether this procedure is sufficiently accurate by studying the radial velocities of galaxies (with respect to the central subhalo) in MTNG. In Fig.~\ref{fig:vrad_prof}, we show the distribution of the normalized radial velocity, $v_{\rm rad}/|v|$, of galaxies, subhaloes, and dark matter particles, at $z = 1$ for the full-physics MTNG run, plotted for the ``high''-density LRGs and ELGs in four mass bins: [12.5, 13.0, 13.5, 14.0] $\pm$ 0.05 in units of $\log(M \ [\hMsun])$. A negative value of $v_{\rm rad}/|v|$ implies inward motion (towards the halo centre), while a positive value implies outward motion. The radial velocity is computed as the relative peculiar velocity between the halo velocity and the object's peculiar velocity, normalized by the relative velocity magnitude.

From the figure, we see that all three types of objects (subhaloes, satellites, and particles) have slightly asymmetric radial motions; i.e.~they are more likely to be moving inwards than outwards, and the asymmetry is most extreme for satellites and subhaloes. We also see evidence for a mass dependence of this behaviour, especially in the satellite distributions. In particular, the LRG asymmetry decreases dramatically with mass: the smallest mass bin is the most extreme, with the majority of satellites being on inward radial orbits rather than moving outwards. This is not surprising given our findings from Fig.~\ref{fig:dist_prof} that LRG satellites in the smallest mass bins live outside the virial radius and are likely on their first infall. The fact that there are much fewer satellites going out suggests that they are likely to be destroyed upon falling into the halo. At the highest mass bins, the survival rate of LRG satellites is highest, which intuitively makes sense as their average concentration is lower (and thus the relative gravitational pull they experience towards the centre is smallest). Our findings for the subhalo population are similar, which emphasizes once again the fragility of infalling subhaloes, specifically in low-mass hosts where the probability of head-on collisions and thus mergers is highest (their concentration, on average, is highest). 

The ELG radial velocity distributions reveal a slightly different story. While we see a similar trend to the LRGs of the asymmetry being exacerbated for lower-mass haloes, it is also noteworthy that the highest mass bin completely inverts that trend. We conjecture that this is the result of quenching, which is strongest in the high-mass haloes. In other words, as an ELG satellite enters a high-mass halo, its star formation rate falls swiftly, quenching the satellite and turning it into a red galaxy in $\lesssim 1 \ {\rm Gyr}$. Thus, the ELG radial velocities are predominantly inwards (see also Fig.~\ref{fig:dist_prof}, which shows that ELGs tend to live on the halo outskirts).

In the case of the particles, their distributions are more symmetric and uniform, though we still see some mass dependence in the asymmetry. In particular, since the particles cannot be destroyed, the asymmetry must be the result of halo dynamics and resolution effects: namely, since haloes are constantly interacting with their surroundings and growing in mass, it is not surprising that there is on average more inbound than outbound structure. If the haloes were perfectly virialized, the particle curves would be more symmetric, though even then, there could be resolution effects breaking the symmetry when particles pass close to the centre of mass \citep{2010MNRAS.402...21N}.

\begin{figure}
\centering  
\includegraphics[width=0.48\textwidth]{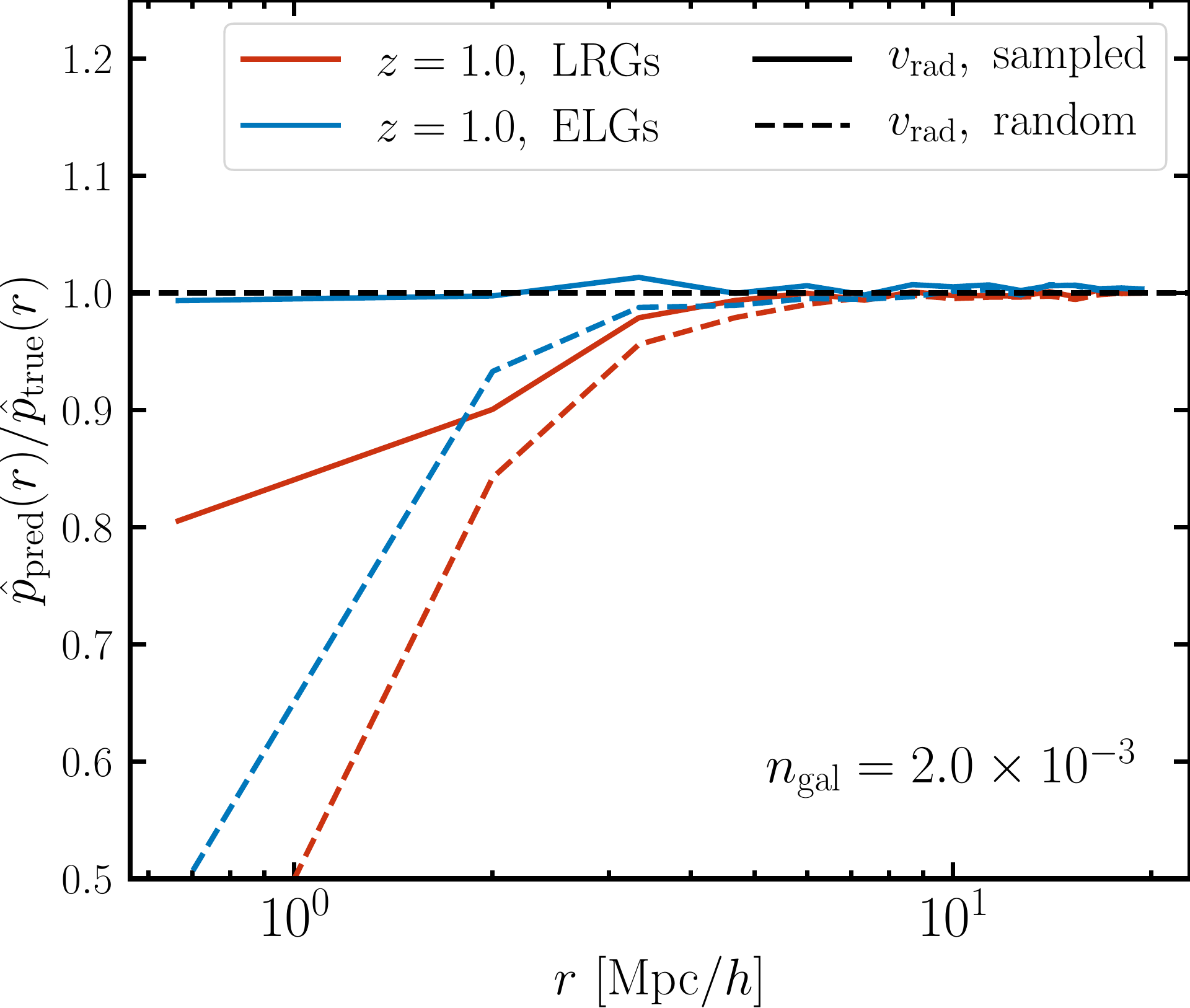}
\caption{Pairwise velocity estimator, $\hat p (r)$, defined as the ratio between the predicted samples and the ``true'' LRG and ELG samples extracted from MTNG at $z = 1$. Here, the predicted samples are obtained by holding the occupations of haloes fixed to their ``true'' values, and only modulating the one-halo population model. In the simpler model (dashed line), we draw randomly from the radial profile and velocity dispersion distributions and choose a random direction for the direction of the vectors for each galaxy, whereas in the second model (solid line), we additionally draw from the radial velocity distributions to determine the radial velocity and then choose the tangential velocity vector randomly. We see that for each of the samples, there is noticeable improvement with respect to the ``truth,'' suggesting that the pairwise velocity statistics is better recovered when adopting a one-halo term that accounts for radial velocity asymmetry. The lingering discrepancy in the LRG case is likely due to the satellite-satellite contribution to $\hat p(r)$, which is not controlled in the $v_{\rm rad}$ sampling model (as $v_{\rm rad}$ is estimated with respect to the centre of the halo). The reason that it plays a larger role in the LRG case is that at large masses the number of satellites per halo is larger (see Fig.~\ref{fig:hod}).}
\label{fig:vrad_corr}
\end{figure}

Having detected the asymmetry in the radial motion of satellites, it is important that we assess how it affects the galaxy correlation statistics. To this end, we devise a simple mechanism that allows us to obtain a mock catalogue with galaxies that have the same average radial velocity profile as the ``true'' satellites in our mocks. Namely, we draw from the normalized radial velocity distributions (Fig.~\ref{fig:vrad_prof}) and the velocity dispersion distribution (Fig.~\ref{fig:disp_prof}) to determine the radial velocity of each satellite and then pick randomly the angle of the tangential velocity. As an idealized test, we compute a modified version of the pairwise velocity estimator \citep{1999ApJ...515L...1F}:
\begin{equation}
    p(r) = \frac{\sum_{ij, \ i < j} ({\hat{\mathbf{r}}}_i -{\hat{\mathbf{r}}}_j) \cdot (\mathbf{v}_i - \mathbf{v}_j)}{\sum_{ij, \ i < j} {1}} \, ,
\end{equation}
where the summations are over all pairs of galaxies $i$ and $j$ (including centrals and satellites), and the final result is binned in separation distance, $\mathbf{r} = \mathbf{r}_i-\mathbf{r}_j$. The denominator corresponds to the data-data pair counts; i.e. ${\rm DD}(r)$. This statistic essentially weights galaxy pairs by their relative velocity along the line that connects them, and as such cannot be computed for real data. It is, nonetheless, helpful for testing whether we recover the correct relative velocities of the galaxies within the halo. The results are shown in Fig.~\ref{fig:vrad_corr}, and we see that for all samples considered there is a substantial improvement in the agreement between MTNG and the mock. 

However, since in observations we rarely have direct measurements of the galaxy velocities, it is more instructive to consider a statistic that indirectly depends on the satellite velocities. Such a statistic is the quadrupole moment of the correlation function, which detects the anisotropy along the line of sight due to redshift space distortions. However, we find that $\xi_{\ell=2}(r)$ is negligibly changed when we switch the radial velocity assignment option on or off (result not shown). This suggests that with the precision that the MTNG simulation provides at this number density, we cannot see a significant difference resulting from the radial velocity asymmetries with this redshift-space statistic.

\subsubsection{Angular distribution}
\label{sec:ang_dist}

\begin{figure}
\centering  
\includegraphics[width=0.48\textwidth]{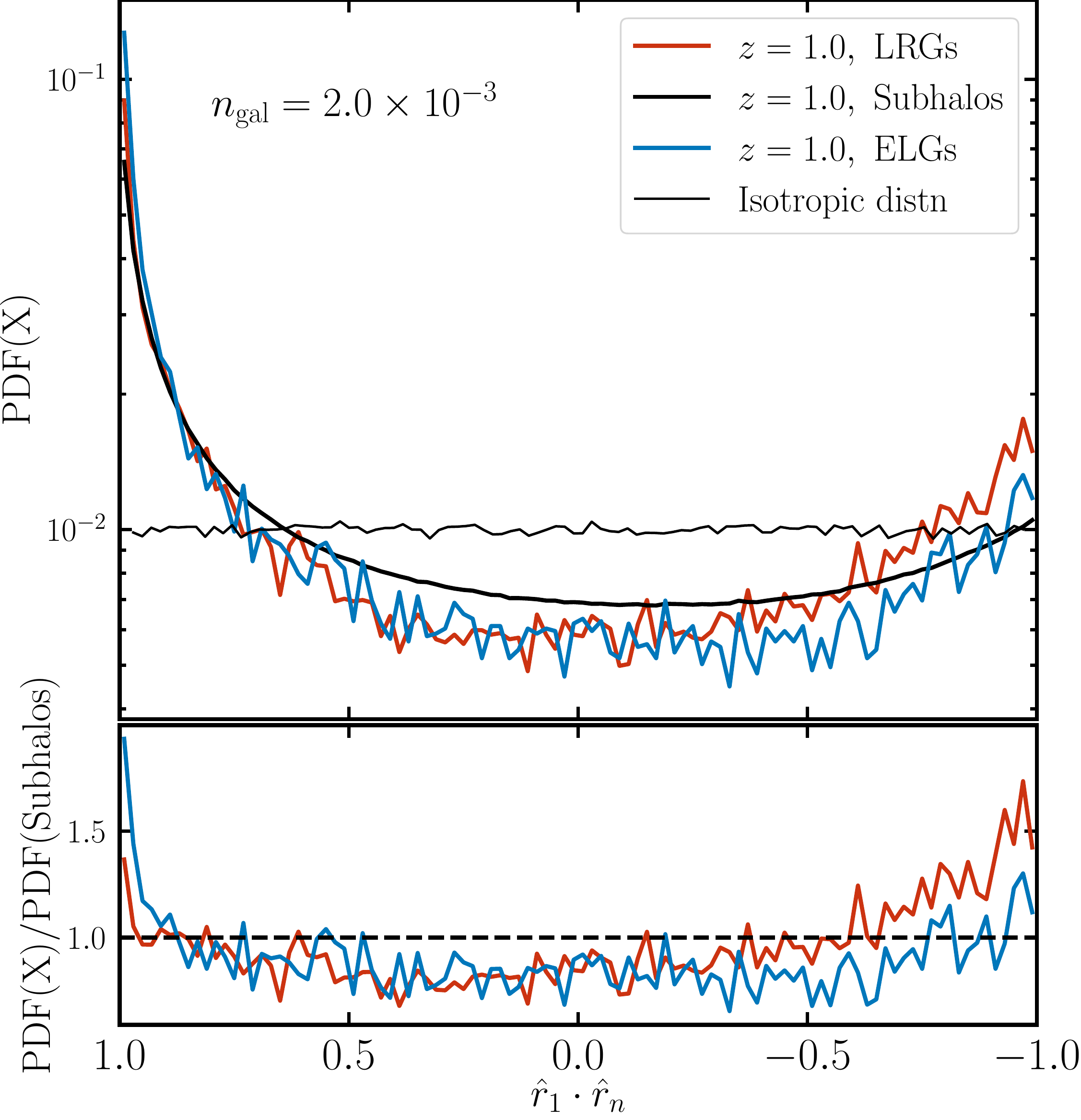}
\caption{Cosine of the angle between the first and the $n^{\rm th}$ satellite ($n > 1$) for the ``high''-density LRG and ELG samples at $z = 1$. The curves are normalized to sum to one. In black, we also show the comparison with a random (isotropic) distribution (thin line) and with all subhalos above $M > 10^9 \hMsun$ (thick line). The bottom panel shows the ratio of the red and the blue PDFs relative to the subhalos. Although both samples show some preference for having the satellites close to each other rather than further apart, we see that the ELGs exhibit more evident anisotropy compared with the LRGs at small angles. In particular, the $n^{\rm th}$ ELG satellite has a 50\% higher probability of being close (within 10$^\circ$) to the first satellite compared with the LRG satellite, and 25\% less chance of being diametrically opposite  ($\sim$ 180$^\circ$) to it. In Fig.~\ref{fig:coop}, we show that this has important consequences for the galaxy clustering. 
}
\label{fig:angle}
\end{figure}

Another intriguing test of the satellite spatial distribution is presented in Fig.~\ref{fig:angle}, where we calculate the cosine of the angle between the first and the $n^{\rm th}$ satellite ($n > 1$) for the ``high''-density LRG and ELG samples at redshift $z = 1$, and compare it with the distribution of all subhalos and a random (isotropic) distribution (the angle between two random vectors is uniform in the range $-1 < \cos(\theta) < 1$). The standard assumption is that satellites are isotropically distributed, which would entail that they follow the thin black curve of Fig.~\ref{fig:angle}. Contrary to that expectation, we find that both samples, as well as the subhalos, show a strong preference for having the satellites subtend a small angle with respect to the largest satellite (or subhalo, respectively). Remarkably, this anisotropy is even more strongly pronounced for the ELGs: compared with the LRGs, the $n^{\rm th}$ ELG satellite is 50\% more likely to be located at an angle smaller than 20$^\circ$ ($\cos(\hat{\mathbf{r}}_1 \cdot \hat{\mathbf{r}}_n) \approx 0.9$) 
with respect to the first satellite, and twice more likely compared with the subhalos. As we show in Fig.~\ref{fig:coop}, this has important consequences for the galaxy clustering of ELGs. Interestingly, the LRGs exhibit a second local maximum at $150^\circ$ ($\cos(\hat{\mathbf{r}}_1 \cdot \hat{\mathbf{r}}_n) \approx -0.9$), suggesting that two LRG satellites have a relatively high chance of being diametrically opposite to each other as well. One potential explanation is that at intermediate angles, i.e. $\sim$90$^\circ$ ($\cos(\hat{\mathbf{r}}_1 \cdot \hat{\mathbf{r}}_n) \approx 0$), the satellites have a higher chance of exerting destructive tidal and gravitational forces onto each other, which is not an issue if they are on opposite sides with respect to the halo centre.

\subsubsection{Proposed satellite distribution model}
\label{sec:sat_prop}

\begin{figure}
\centering  
\includegraphics[width=0.48\textwidth]{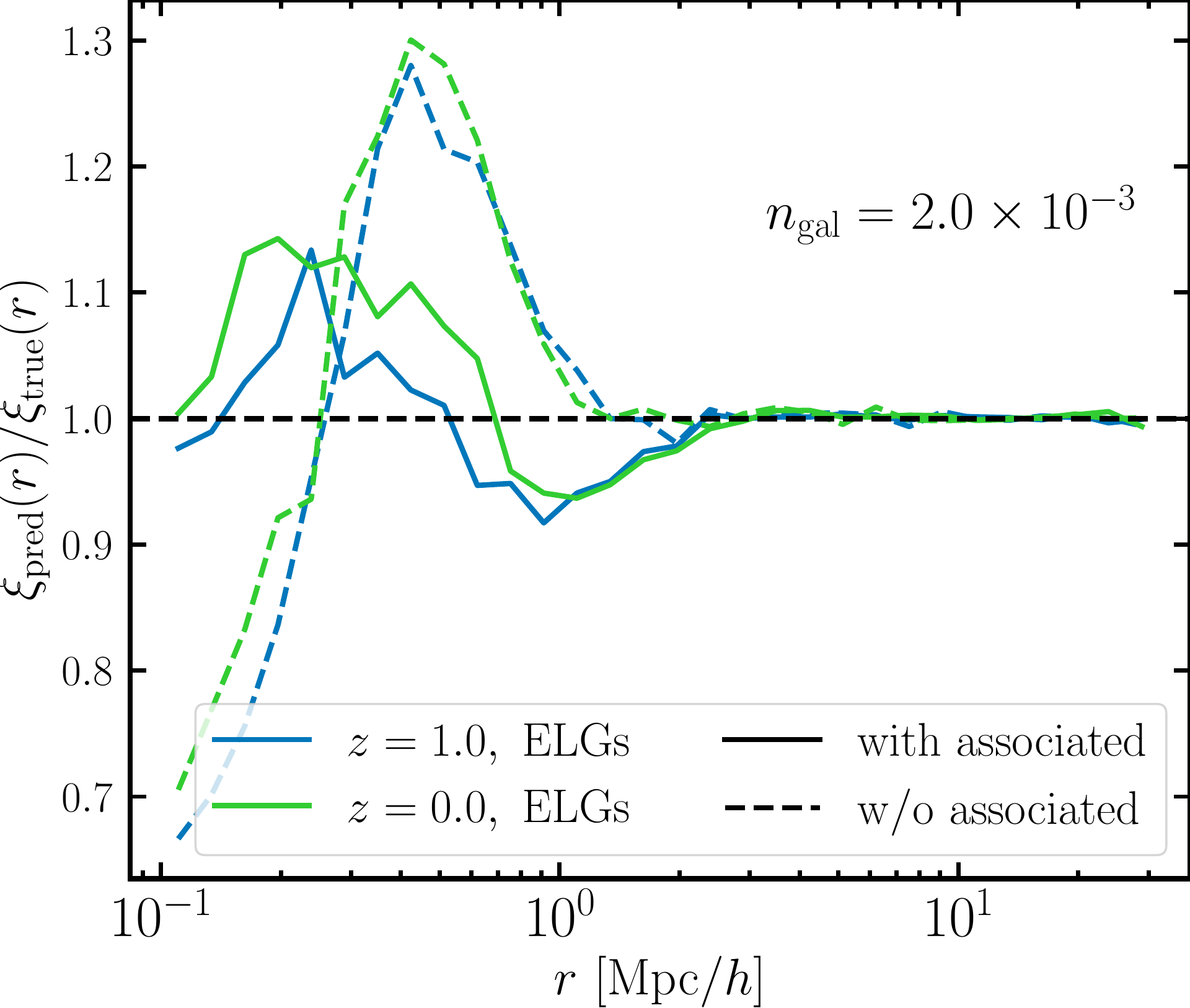}
\caption{Correlation function ratio between the predicted samples and the ``true'' high-density ELG samples extracted from MTNG at $z = 1$ and $z=0$. Here, the predicted samples are obtained by holding the occupations of haloes fixed to their ``true'' values and only modulating the one-halo population model. In the simpler model (dashed line), we draw randomly from the radial profile and velocity dispersion distributions and choose a random direction for the radial and velocity vectors for each galaxy, whereas in the second model (solid line), we apply the ``associated'' satellite model detailed in the text: with probability of $X_{\rm assoc.}$, each satellite is either ``associated'' with the first satellite or the central subhalo; if ``associated'', its radial distance from the first satellite is drawn from a Gaussian distribution with $(\mu_{\rm assoc.}, \ \sigma_{\rm assoc.})$; otherwise, we draw from the true radial distributions (Fig.~\ref{fig:dist_prof}). We see that the standard model; i.e. ``w/o associated'', shows a significant discrepancy in the range $0.1\, \hMpc < r < 1 \, \hMpc$, despite matching the radial profiles of the ``true'' ELG sample, conveying strong hints of the anisotropic distribution of ELG satellites (confirmed in Fig.~\ref{fig:angle}). The discrepancy is reduced by more than 20\% when we adopt the ``associated'' model. Note that the value of $X_{\rm assoc.}$ could be further adjusted to match the one-halo term of MTNG even better.}
\label{fig:coop}
\end{figure}

Once the central and satellite occupations have been assigned to each halo via some recipe \citep[see also section~3.1 of our companion paper,][for our improved assignment model]{Hadzhiyska2022b}, we distribute the satellites in each halo according to the following procedure. At fixed halo mass (i.e., in thin bins of equal mass), we: 
\begin{enumerate}
    \item Draw samples from the ``true'' satellite radial distribution at that mass bin to determine the distances of the mock satellites from their respective halo centres. The angles of the position vectors are picked randomly.  
    \textit{Exception:} In the case of ELGs, we notice that the satellites are more likely to come in close-knit groups (see Fig.~\ref{fig:angle}), so we adopt a more complex procedure: namely, we assign a fractional chance $X_{\rm assoc.}$ that the $n^{\rm th}$ ($n > 1$) ELG satellite that appears in a host halo is ``associated'' with the first satellite rather than the central \citep[which could be explained through the idea of cooperative galaxy formation][]{1993ApJ...405..403B}. If the satellite is ``associated,'' then we draw from a Gaussian with mean $\mu_{\rm assoc.}$ and width $\sigma_{\rm assoc.}$ in a random direction centred on \textit{the first satellite}. Otherwise, we proceed as usual and draw from the ``true'' radial distribution around the central. Empirically, we set $X_{\rm assoc.} = 50\%$ and $(\mu_{\rm assoc.}, \ \sigma_{\rm assoc.})=(0.1, 0.1)$. Fig.~\ref{fig:coop} shows the effect that switching this ELG ``association'' procedure on and off has on the galaxy clustering. We see that despite the fact that in both scenarios we have matched the radial profiles of the ``true'' ELGs, it is only when we turn on the ``associations'' that we can reconcile the clustering discrepancy on small scales. This has important implications for halo modelling as it suggests that ELGs exhibit non-isotropic or cooperative behaviour. We find this to be much less the case for the LRGs.
    \item Draw samples from the ``true'' satellite velocity dispersion distribution at that mass bin to determine the magnitude of the velocity. The angles of the velocity vectors are either picked randomly or the normalized radial velocities are drawn from the ``true'' distribution (i.e. Fig.~\ref{fig:vrad_prof}) and the tangential velocities are picked randomly.  
\end{enumerate}

Note that the free parameters of this model, $X_{\rm assoc.}$, $\mu_{\rm assoc.}$ and $\sigma_{\rm assoc.}$ can be adjusted to improve the clustering agreement, if needed.

\section{Conclusions}
\label{sec:conc}

In this work, we propose a satellite population model that challenges standard assumptions of the galaxy-halo connection and improves the agreement with state-of-the-art hydrodynamical simulations by introducing minimal and flexible modifications to the standard approach. The main application of our model lies in the analysis of large-scale galaxy surveys with the help of HOD-like techniques applied to simulated halo catalogues. We incorporate several optional decorations of the HOD approach and discuss in detail their effects on the galaxy clustering. Due to the modular nature of these decorations, our findings are easily translatable to models that make use of the particle and subhalo catalogues, and can thus be helpful in capturing the galaxy-matter correlation. This work does not assume that the MTNG model reproduces the complex physics of the real Universe perfectly, but rather it should be viewed as an illustrative example of which assumptions might reasonably be falsified and subsequently be improved in galaxy-halo modeling.

Our conclusions regarding the one-halo term can be summarized as follows:
\begin{itemize}
    \item When studying the occupation statistics of ELG satellites, we found evidence of super-Poissonian behaviour, challenging the traditional assumption made by HOD models that the satellite occupations are Poisson distributed (see Fig.~\ref{fig:hod}). We then designed a simple recipe in Section~\ref{sec:non_poiss} that allows us to populate haloes with satellites according to a modified pseudo-Poisson distribution, at the cost of adding a single free parameter to the model. We showed in Fig.~\ref{fig:pseudo} that this correction is able to improve the one-halo term predictions for ELGs.
    \item Furthermore, we discovered that the central and satellite occupations of ELGs are correlated (see Fig.~\ref{fig:prob_cen}), in disagreement with the common notion that the probability of a halo hosting a central is independent of its probability to host a satellite. We offered a simple solution in Section~\ref{sec:cond_prob} that requires the inclusion of only one additional parameter for controlling the conditional probability of central occupation, and we demonstrated in Fig.~\ref{fig:cond_prob} that this change is capable of significantly reducing the discrepancy. 
    \item Another surprising finding about the ELG satellite distribution is that compared with LRGs, they appear to be significantly more anisotropically arranged within the halo. Through Fig.~\ref{fig:angle}, we showed that ELG satellites are often paired (or ``associated'') with one another, so that assuming spherical symmetry leads to a poor prediction of the one-halo term. We propose a model that gives non-zero probability to satellites being more strongly ``associated'' with other satellites rather than the centre of the halo, which helps to improve the agreement in the one-halo regime (see Fig.~\ref{fig:coop}).
    \item We argue that the three findings we list above provide strong evidence for cooperative galaxy formation (or galaxy conformity) that predominantly affects the ELG population. We conjecture that star formation may be triggered on large scales from the collapse of density peaks, which leads to the formation of a population of blue (ELG) galaxies. After a while ($\sim$1 Gyr), the galaxies drift apart from each other and turn into red galaxies (LRGs) due to quenching.
    \item The galaxy assignment procedure we adopt follows the satellite profiles and velocity dispersion distributions of the true galaxies at fixed mass (if using the extended halo model, also at fixed secondary and tertiary halo property), shown in Fig.~\ref{fig:dist_prof} and Fig.~\ref{fig:disp_prof}. We find substantial evidence of FoF percolation; i.e. the chaining of two or more dark-matter haloes strung together into a single massive FoF group which is treated by standard HOD models as a single halo. FoF percolation manifests itself in the satellite profiles of low- and intermediate-mass haloes by contributing a large number of satellites outside the virial radius. We plan to further investigate this issue, which is also relevant to the question of assembly bias, with the spherical-overdensity-based halo finder \textsc{CompaSO} \citep{2022MNRAS.509..501H} in a future study.  
    \item We also found that the radial velocities of galaxies are strongly skewed towards the direction of infall (Fig.~\ref{fig:vrad_prof}). We propose a model that assigns radial velocities to the mock galaxies following the true distribution (while randomizing the tangential component). However, when compared with a model assuming random velocity directions, we do not find a signature of this effect in the redshift space clustering via the quadrupole, $\xi_{\ell=2}$, with the current level of precision. 
    \item In all of the above comparisons (Figs.~\ref{fig:dist_prof}-\ref{fig:vrad_prof}), we find that neither the particle nor the subhalo catalogues follow the ``true'' galaxy distribution out of the box, but when equipped with the proper decorations, the one-halo term can be substantially improved.
\end{itemize}

With every step forward taken by observational surveys, numerical studies must follow by maximizing their accuracy and applicability to galaxy observations. In this finely tuned harmony, the next generation of simulations must be equipped with optimal galaxy-halo models that are capable of reliably reproducing galaxy observables at sufficiently large scales. Analyses refining and further developing these models with the help of state-of-the-art hydrodynamical simulations are thus invaluable for increasing the cosmological yield of large-scale structure surveys, and for attaining a profound understanding of the physics that governs our Universe.

\section*{Acknowledgements}

The authors gratefully acknowledge the Gauss Centre for Supercomputing (GCS) for providing computing time on the GCS Supercomputer SuperMUC-NG at the Leibniz Supercomputing Centre (LRZ) in Garching, Germany, under project pn34mo. This work used the DiRAC@Durham facility managed by the Institute for Computational Cosmology on behalf of the STFC DiRAC HPC Facility, with equipment funded by BEIS capital funding via STFC capital grants ST/K00042X/1, ST/P002293/1, ST/R002371/1 and ST/S002502/1, Durham University and STFC operations grant ST/R000832/1. CH-A acknowledges support from the Excellence Cluster ORIGINS which is funded by the Deutsche Forschungsgemeinschaft (DFG, German Research Foundation) under Germany’s Excellence Strategy – EXC-2094 – 390783311. VS and LH acknowledge support by the Simons Collaboration on “Learning the Universe”. LH is supported by NSF grant AST-1815978. SB is supported by the UK Research and Innovation (UKRI) Future Leaders Fellowship [grant number MR/V023381/1]. SC acknowledges the support of the ``Juan de la Cierva Incorporac\'ion'' fellowship (IJC2020-045705-I).


\section*{Data Availability}

The data underlying this article will be shared upon reasonable request to the corresponding 
authors. All MTNG simulations will be made publicly available in 2024.



\bibliographystyle{mnras}
\bibliography{example} 








\bsp	
\label{lastpage}
\end{document}